\documentclass[twocolumn]{aastex62}
\pdfoutput=1 
\usepackage{amssymb}
\usepackage{graphicx}
\usepackage{amsmath}
\usepackage{pbox}
\usepackage{xspace}
\usepackage{array}
\usepackage{apjfonts} 
\usepackage{fixltx2e}
\usepackage{color}

\shorttitle{Luminous Transient Light Curves at Peak}
\shortauthors{Khatami and Kasen}

\begin{document}

\title{Physics of Luminous Transient Light Curves: A New Relation Between Peak Time and Luminosity}

\author{David K. Khatami}
\affiliation{Department of Astronomy, University of California, Berkeley, CA, 94720}
\author{Daniel N. Kasen}
\affiliation{Department of Astronomy, University of California, Berkeley, CA, 94720}
\affiliation{Nuclear Science Division, Lawrence Berkeley National Laboratory, 1 Cyclotron Road, Berkeley, CA 94720}

\begin{abstract}
    Simplified analytic methods are frequently used to model the light curves of supernovae and other energetic transients and to extract physical quantities, such as the ejecta mass and amount of radioactive heating. The applicability and quantitative accuracy of these models, however, have not been clearly delineated. Here we carry out a systematic study comparing certain analytic models to numerical radiation transport calculations. We show that the neglect of time-dependent diffusion limits the accuracy of common Arnett-like analytic models, and that the widely-applied Arnett's rule for inferring radioactive mass does not hold in general, with an error that increases for models with longer diffusion times or more centralized heating. We present new analytic relations that accurately relate the peak time and luminosity of an observed light curve to the physical ejecta and heating parameters. We further show that recombination and the spatial distribution of heating modify the peak of the light curve and that these effects can be accounted for by varying a single dimensionless parameter in the new relations. The results presented should be useful for estimating the physical properties of a wide variety of transient phenomena.
\end{abstract}

\section{Introduction}
\label{sec:intro}

Wide-field surveys are gathering data on an increasing number of common supernovae (SNe)  and related transients such as  tidal disruption events, fast-evolving luminous transients,
superluminous supernovae, and kilonovae.
The general physics controlling the light curves of these events is similar -- energy deposited either by a propagating shock (e.g. Type II-P SNe) or a heating source (e.g. radioactivity or a central engine) radiatively diffuses through the optically thick and expanding ejecta, undergoing adiabatic loses until the radiation reaches the surface and escapes. Analysis of the observed light curves can provide information on the ejecta properties and the nature of the powering source.
 
With the increasing number of observed transients, there is increased need for fast, empirical techniques to infer their physical properties and discriminate between competing theoretical explanations.
Simplified analytic models are commonly used to analyze observations and make theoretical predictions (e.g. \cite{Li1998TransientMergers,Chatzopoulos2012,Inserra2013,Villar2017,Nicholl2017,Guillochon2018}). Most common among these are the Arnett models \citep{Arnett1980,Arnett1982}, in which  bolometric light curves are calculated through a simple numerical integral. These models also provide several ``rules of thumb'' for estimating physical properties from the light curve brightness and duration.  In particular, ``Arnett's rule'' states that the instantaneous heating rate at peak is  equal to the peak luminosity \citep{Arnett1982}. For Type I supernovae, this rule in principle allows one to extract the mass of radioactive $^{56}$Ni \citep{Stritzinger2006,Valenti2007,Drout2011,Scalzo2014,Prentice2016}.

Despite the frequent application of these analytic models and rules, a systematic study of their accuracy and applicability has not been carried out. Previous numerical models of Type Ia SNe have noted that Arnett's rule is usually accurate to $\sim 20\%$ \citep{Blondin2013,Hoeflich2017LightObservations,Khokhlov1993LightMechanisms}. 
For Type Ib/c SNe Arnett's rule is typically off by $\sim 50\%$ \citep{Dessart2015,Dessart2016}, and for Type II SNe like SN1987A the error is a factor of $\sim 2$  \citep{Woosley1988}. The reasons for these discrepancies -- and why they are more extreme for certain classes of transients -- have not been fully spelled out.

In this paper we carry out a systematic investigation of certain analytic models compared to numerical light curve simulations. We find that the main limitations in the Arnett models stem from the assumption that a self-similar temperature profile is immediately established in the ejecta. This fails to account for the time-evolving propagation of a radiative ``diffusion wave'' from the heating source to the surface. The neglect of the diffusion wave is worse for more centrally concentrated heating sources, which is why Arnett's rule is worse for more stratified Type~Ib/c SNe and better for more thoroughly mixed Type Ia SNe.

We derive a new relation between the peak time and peak luminosity of transient light curves which accurately captures the results of numerical models. We  study how the relation depends on the spatial distribution of heating as well as the effects of a non-constant opacity due to recombination. The new relation is parameterized by a dimensionless constant $\beta$, and works for a variety of  assumed heating sources and ejecta characteristics.

In Section 2, we describe the assumptions and limitations of  the Arnett light curve models and Arnett's rule. In Section 3, we derive the new peak-time luminosity relation and compare it to numerical simulations. In Section 4, we investigate the relation between the peak and diffusion time. In Section 5, we examine the effects of concentration of the heating source. In Section 6, we look into the effects of a non-constant opacity due to recombination on the light curve and the new relation. Finally, in Section 7, we apply the results and relation to radioactive $^{56}$Ni-powered transients. In Appendix A, we provide a table of the new relation for a variety of luminous transients.

\section{Limitations of Arnett-like Models}

The analytic light curve modeling approach of \cite{Arnett1980,Arnett1982} (hereafter A80,A82) is widely used to analyze luminous transients. A closely related ``one-zone'' modeling approach \citep{Arnett1979,Kasen2010,Villar2017} differs in its mathematical details but results in a similar expression for calculating the light curve. 

\subsection{Assumptions and Light Curve Solution}

The Arnett-like models begin with the first law of thermodynamics 
\begin{align}\label{eqn:firstlaw}
\dot{\mathcal{E}}= - P\dot{\mathcal{V}} + \varepsilon-\frac{\partial L}{\partial m}
\end{align}
where $\mathcal{E}$ is the specific (i.e. per unit mass) energy density, $P$ the pressure, $\mathcal{V}=1/\rho$ is the specific volume,  $\varepsilon$ is the specific heating rate, and $L$ the emergent luminosity. Several simplifying assumptions are then made:
(1) the ejecta is expanding homologously and so the radius evolves as
\begin{align}
    R_{\rm ej}(t)=v_{\rm ej}t
\end{align}
where $v_{\rm ej}$ is the maximum ejecta velocity; (2) Radiation pressure dominates over the gas pressure and so we can express the specific energy density as
\begin{align}
\mathcal{E}=3P/\rho=aT^4/\rho
\end{align}
where $T$ is the temperature; (3) The luminosity is described by the spherical diffusion equation
\begin{align}\label{eqn:diffusion}
L(r) =- 4 \pi r^2 \frac{c}{3\kappa\rho}\frac{\partial e}{\partial r}
\end{align}
where $e=\rho\mathcal{E}$ is the energy density (per unit volume) and $\kappa$ the opacity; and (4) the ejecta is characterized by a constant opacity.

The Arnett models make an additional consequential, but often overlooked, assumption: (5)
The energy density profile is self-similar, i.e., the spatial dependence is fixed and only the overall normalization changes with time

\begin{align}
e(x,t) = \frac{E_{\rm int}(t)}{V(t)} \psi(x)
\label{eq:ss_profile}
\end{align}
where $E_{\rm int}$ is the total internal energy of the ejecta
and $x=r/R_{\rm ej}(t)$ is the (comoving) dimensionless coordinate. The dimensionless function
$\psi(x)$ describes the spatial dependence of the radiation energy density, which by assumption does not change with time. 
Substituting Eq.~\ref{eq:ss_profile} into the diffusion equation (Eq.~\ref{eqn:diffusion}) gives the emergent luminosity
at $r = R_{\rm ej}$
\begin{align}
L = \frac{t E_{\rm int}(t)}{\tau_d^2}
\label{eq:L_Eint}
\end{align}
where 
\begin{align}
\tau_d= \left[ \frac{3}{4 \pi} 
\frac{\kappa M_{\rm ej}}{v_{\rm ej}c} \frac{1}{\xi} \right]^{1/2}
\end{align}
is the characteristic diffusion time through the ejecta with mass $M_{\rm ej}$. The quantity $\xi = d \psi/dx\vert_{x=1}$ specifies
the energy density gradient at the ejecta surface and is a constant when self-similarity
is assumed. The one-zone models  make the {\it ansatz} $\xi = 1$.
\cite{Arnett1982} uses a more sophisticated separation of variables method to 
derive a self-consistent solution for $e(x,t)$. This   requires making a final assumption: (6) The spatial distribution of the heating is proportional to the energy density. Eq.~\ref{eqn:firstlaw} can then be solved to find
\begin{align}\label{eqn:arnett_egy}
e(x,t) = \frac{E_{\rm int}(t)}{V(t)} 
\left[ \frac{\pi}{3} \frac{\sin (\pi x)}{x}\right]
\end{align}
which gives $\xi = \pi^2/3$.

To solve for the light curve, the Arnett models integrate Eq.~\ref{eqn:firstlaw}
over the entire ejecta and apply the assumptions of homology and radiation energy domination
to derive an equation for global energy conservation
\begin{align}
\frac{ d E_{\rm int}(t) }{d t} = -  \frac{E_{\rm int}(t)}{t}
+ L_{\rm heat}(t) - L(t)
\label{eq:global_energy}
\end{align}
where  $L_{\rm heat}(t)$ is the total input heating rate. Using Eq.~\ref{eq:L_Eint} to replace $E_{\rm int} = L \tau_d^2/t$ and rearranging gives
\begin{align}
\frac{\tau_d^2}{t} \frac{ d L}{d t} =   L_{\rm heat}(t) - L(t)
\label{eq:global_energy_L}
\end{align}
Arnett's rule follows, since the condition of an extremum $d L/dt = 0$ implies  $L = L_{\rm heat}$.

The ordinary differential equation Eq.~\ref{eq:global_energy_L} can be solved for $E_{\rm int}(t)$ and hence the emergent luminosity
\begin{align}\label{eqn:arnett}
L(t)=\frac{2}{\tau_d^2}e^{-t^2/\tau_d^2}\int_0^t\,t' L_{\rm heat}(t')e^{t'^2/\tau_d^2}\,{\rm d}t'
\end{align}
Both the one-zone  and the separation of variables approaches result in the same expression for the light curve; the only difference is the value of $\xi$, which reflects different assumptions about the shape of the self-similar energy density profiles. The diffusion time for the one-zone models is  a factor of $\pi/\sqrt{3} \approx 2$ larger.
To avoid confusion, we hereafter define a characteristic diffusion timescale without any numerical factors
\begin{align}
t_d=\sqrt{\frac{\kappa M_{\rm ej}}{v_{\rm ej}c}}
\end{align}
and so $\tau_d = [ 3/ 4 \pi \xi]^{1/2} t_d$.
Physically, the characteristic diffusion timescale gives the time at which the expansion timescale $t_{\rm exp}=R_{\rm ej}/v_{\rm ej}$ equals the diffusion time $t_{\rm diff}=\kappa\rho R_{\rm ej}^2/c$. The peak time scales with the diffusion timescale $t_{\rm peak}\propto t_d$, but the numerical coefficient relating them depends on the distribution of heating, nature of the opacity, and other effects.

The self-similarity assumption will be shown below to limit the accuracy of the Arnett models. However, this is not a necessary assumption as Eq.\ref{eqn:arnett_egy} is only the first eigenfunction of the separated spatial equation.
The full solution of the energy density can be expressed as an infinite sum of higher order eigenfunctions whose normalization will be set by the spatial distribution of heating and boundary conditions. \cite{Pinto2000} show how such an approach can be used to
relax the assumptions (5) and (6) and produce more accurate light curves. However, due to the more complicated nature of the solution, the full solution with higher-order eigenmodes is rarely used in practice.

\begin{figure}
    \centering
    \includegraphics[width=0.5\textwidth]{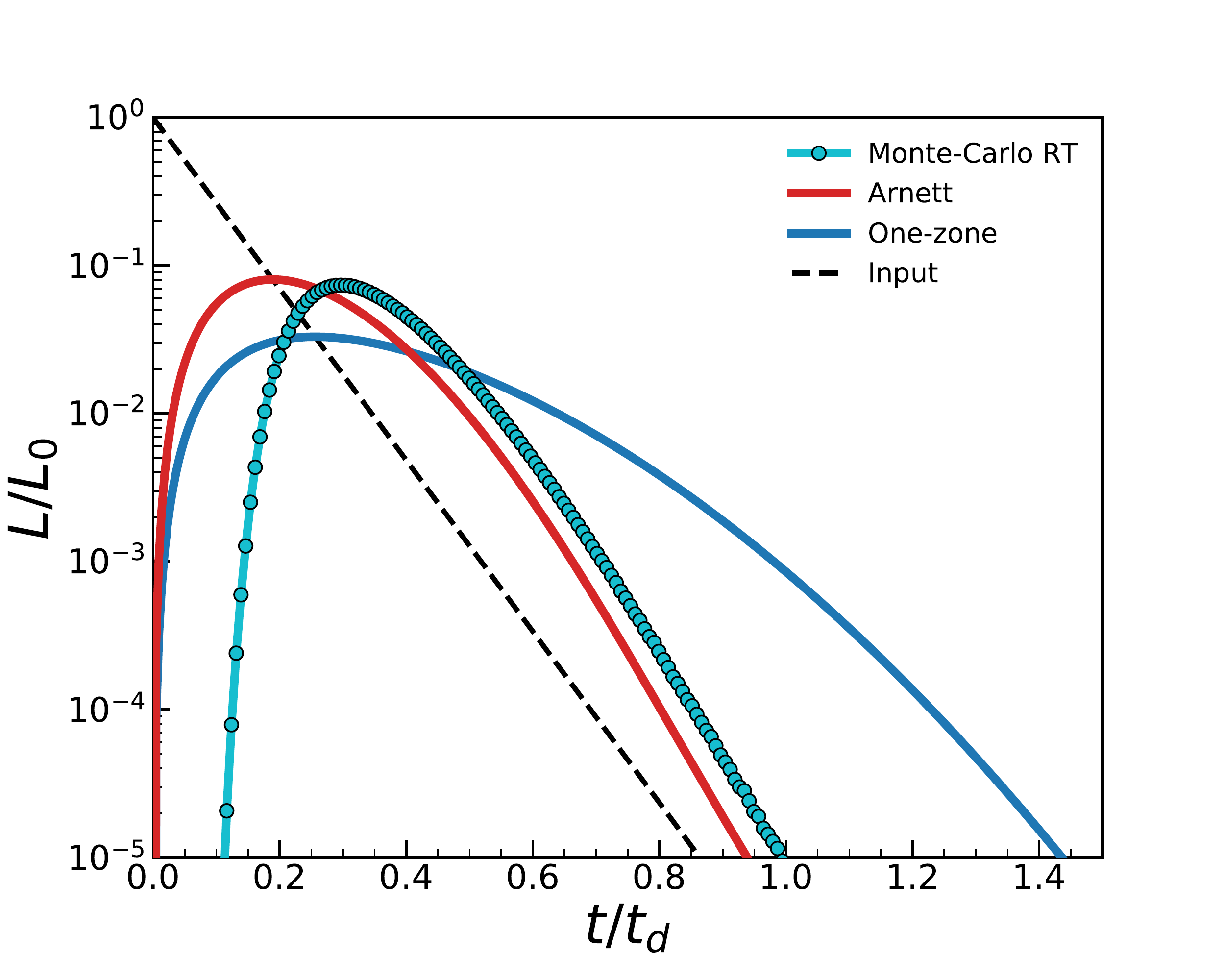}
    \caption{Light curves from the Arnett solution Eq.(\ref{eqn:arnett}) with different choice in the diffusion timescale factor $\xi$ (red and blue lines), compared with a numerical monte-carlo radiation transport solution using Sedona (teal line with points). The input heating (dashed black line) consists of a centrally-located exponential source with luminosity $L_{\rm in}(t)=L_0\exp\left[-t/t_s\right]$ with a timescale $t_s=10$ days and a characteristic diffusion timescale $t_d=100$ days.}
    \label{fig:sedona_arnett_comp}
\end{figure}

\subsection{Comparison to Numerical Simulations}
To assess the accuracy of the Arnett solutions, we compare them to numerical monte-carlo radiation transport calculations run with Sedona \citep{kasen2006}. We adopt similar assumptions as A82: homologous expansion, uniform density, and a constant opacity. Non-constant opacity will be considered in Section 6. In this section, the ejecta has a diffusion timescale $t_d = 100$~days and the heating source is at the center and follows
$L_{\rm heat}(t) = L_0 e^{-t/t_s}$, where the timescale $t_s=10$ days.

Fig. (\ref{fig:sedona_arnett_comp})  compares the numerical light curve to the Arnett analytic solution. 
The numerical models have an initial ``dark period'' until $t\sim 0.1t_d$, before which the photons have not had sufficient time to diffuse from the center of the ejecta \citep{Piro2013}. 
In contrast, the analytic solutions predict a steeper rise beginning at $t=0$, a consequence of the assumption that radiation energy is immediately distributed throughout the ejecta.  
The A82 solution predicts a peak time a factor of 2 shorter than the numerical result, but gives roughly the correct peak luminosity. The peak time of the one-zone model is closer to the numerical simulation, but under-predicts the peak luminosity and is overall too broad. There is no choice of $\xi$ such that the analytic solution closely matches the numerical light curve.

\begin{figure}
    \centering
    \includegraphics[width=0.5\textwidth]{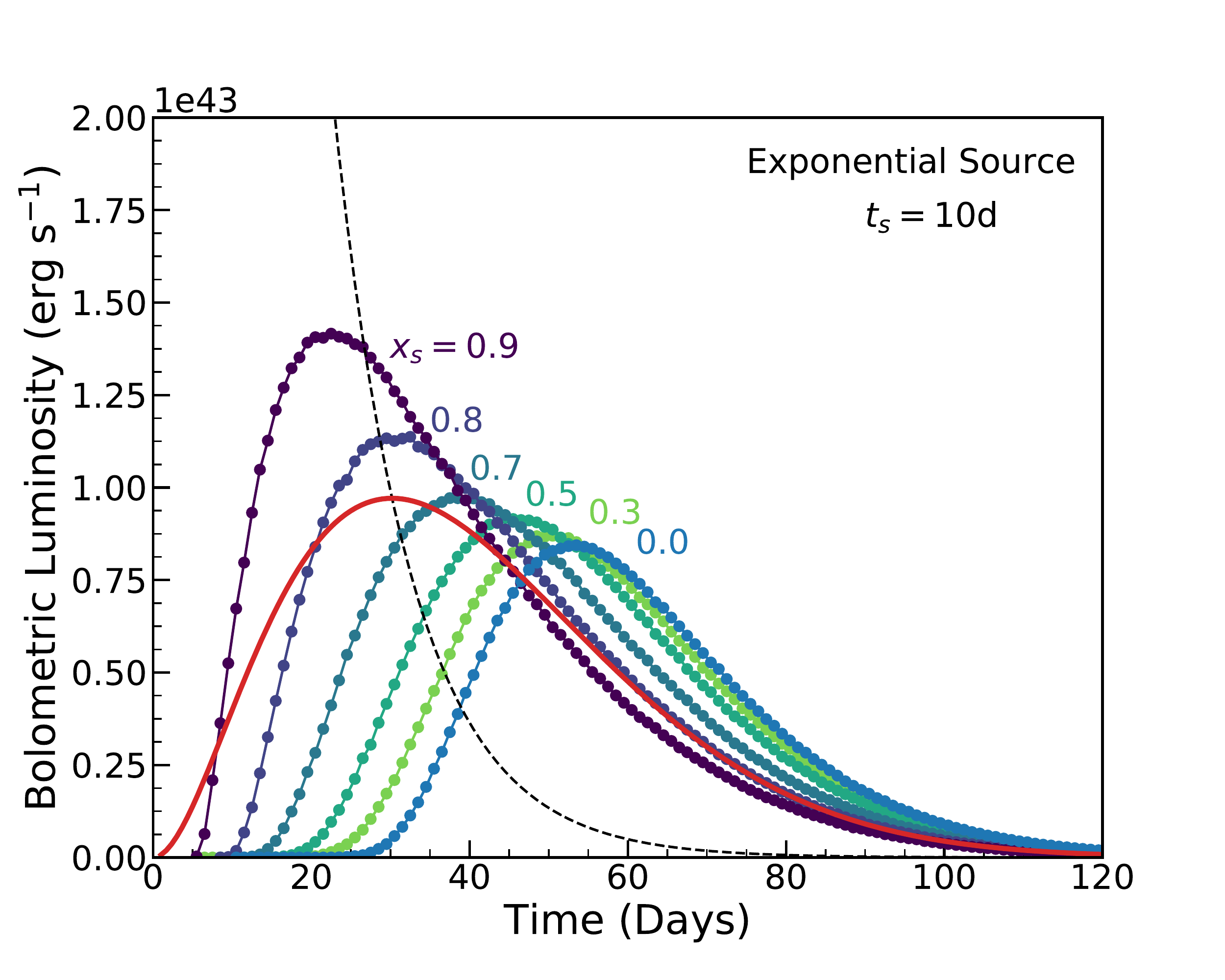}
    \caption{Effects of varying concentration, in terms of the dimensionless radius $x_s$. The heating is uniformly mixed out to $x_s$. The solution of A82 is shown for comparison (red line), as well as the input heating rate (dashed black line), here an exponential source with timescale $t_s=10$ days.
    Arnett's rule, which predicts that the input heating rate should intersect the observed light curve exactly at peak holds only for well mixed sources,
    $x_s \approx 0.8$.
    }
    \label{fig:exp_mix_lcs}
\end{figure}

This inaccuracy of the analytic models is more pronounced for more centrally
concentrated heating  sources. Fig. (\ref{fig:exp_mix_lcs}) shows numerical models where the heating source has been uniformly mixed out to dimensionless radius $x_s$, with $x_s=0$ corresponding to a central source. 
The Arnett analytic solution most closely resembles a well-mixed numerical model with $x_s \approx 0.8$.
We can define a heating-weighted radius where the bulk of heating occurs as
\begin{align}
    \langle x_s\rangle = \left(\frac{\int_0^1 x^2 \dot{e}_{\rm heat}(x)\,dx}{\int_0^1 \dot{e}_{\rm heat}(x)\,dx}\right)^{1/2}
\end{align}
where $\dot{e}_{\rm heat}(x)$ is the energy density heating rate at $x$. For constant heating out to radius $x_s$, we have the relation $\langle x_s\rangle = x_s/\sqrt{3}$.
In the Arnett solution,  $\dot{e}_{\rm heat}(x)\propto e(x)$ and using Eq.(\ref{eqn:arnett_egy}) we find that  $\langle x_s\rangle\approx 0.4$, which indeed corresponds to $x_s \approx 0.7$.

\begin{figure}
    \centering
    \includegraphics[width=0.5\textwidth]{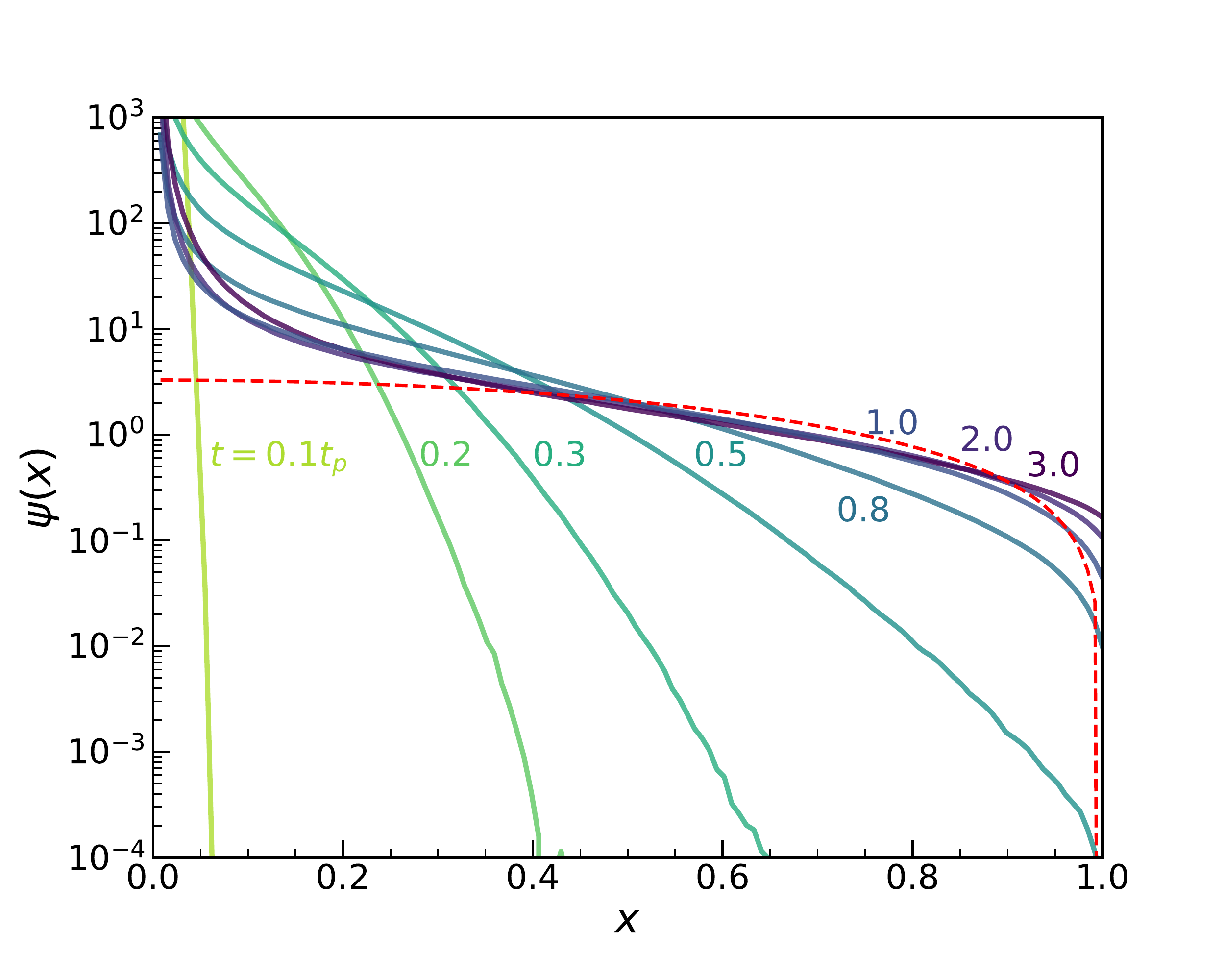}
    \caption{Evolution of the energy density profile Eq. (\ref{eq:ss_profile}) for a central exponential heating source. Shown are profiles at different times relative to peak, $t_p$. The self-similar assumption breaks down at early times as the diffusion wave propagates outward. At times $t>t_p$, the profile settles into a self-similar shape. Also shown for comparison is the solution of Arnett in Eq.(\ref{eqn:arnett_egy}) (dashed red line).}
    \label{fig:selfsimilar}
\end{figure}

The limitation of the Arnett models stems from the assumption that the spatial distribution of the radiation field is self-similar. In reality, for central sources a radiation diffusion wave initially propagates outwards, only reaching the surface and establishing a self-similar profile after a timescale $\sim t_d$.  In Fig. (\ref{fig:selfsimilar}), we show the evolution of the energy density profile defined in Eq.(\ref{eqn:arnett_egy}) for a central exponential heating source. At early times, a diffusion wave propagates outwards, and the self-similar assumption fails.
By neglecting this diffusion wave, the Arnett-like models overestimate the luminosity at early times. For a more uniformly mixed source, self similarity is established earlier, and so the Arnett models are more applicable.

Thus, it is a common misconception that the Arnett models assume a centrally-located heating source --  while the \textit{energy density} increases towards the center, the \textit{heating luminosity} peaks close to the surface, producing the faster rise and earlier peak compared to the central-heating numerical solution. 


This likely explains why the A82 solution more closely predicts Type~Ia rather than core collapse SN light curves  -- Type~Ia SN typically have a much larger degree of mixing \citep{Blondin2013} while core collapse SNe have more centrally concentrated $^{56}$Ni. We explore the effects of the spatial distribution of the heating source in more detail in Section 5.

\subsection{Arnett's Rule}

A specific prediction of the Arnett models  is that the peak luminosity is equal to the heating rate at peak, i.e.
\begin{align}
    L_{\rm peak}=L_{\rm heat}(t_{\rm peak})
\end{align}
This is commonly referred to as Arnett's rule (or Arnett's law) and is widely used to infer e.g. the nickel mass in radioactive SNe \citep{Stritzinger2006,Valenti2007}.

We see from Fig.(\ref{fig:exp_mix_lcs}) that Arnett's rule does not hold in general and that its accuracy depends on the heating source concentration. For centrally concentrated sources ($x_s \lesssim 0.8$) Arnett's law gives an {\it under-estimate} of the true peak luminosity, with the error being systematically worse for more centralized heating.  For nearly fully mixed heating sources ($x_s \gtrsim 0.8$) Arnett's law is an over-estimate.

The failure of Arnett's rule again stems from the assumption of self-similarity,  which implies a proportionality between the luminosity and the total ejecta internal energy, $L \propto E t$. Under this assumption, the light curve has to peak simultaneously with $E(t)$ -- i.e.,  at the time  when the rate of energy loss, $L(t)$ equals the rate of energy gain, $L_{\rm heat}(t)$. In reality, the time-dependent propagation of a diffusion wave means that the luminosity does not strictly track the internal energy. For central sources, $L(t)$ generally lags $E(t)$ and the light curve peaks at a time when $L > L_{\rm heat}$. For fully mixed cases, $L(t)$ leads $E(t)$ and the light curve peaks when $L < L_{\rm heat}$.  Arnett's law holds only for the case of a specific concentration ($x_s \approx 0.8$) for which the light curve coincidentally peaks at the same time as does the internal energy.

\section{A New Relation Between Peak Time and Luminosity}

Given the limits of Arnett's rule, we look for a more robust relationship between the peak time
and peak luminosity of a transient light curve. We proceed by  considering the evolution of the global internal energy, $E$,  and rewrite Eq.~\ref{eq:global_energy} as
\begin{align}
\frac{d(t E )}{dt}= t [L_{\rm heat}(t)-L(t)]
\end{align}
which integrates to
\begin{align}\label{eqn:integ_relation}
tE(t) = \int_0^t t' L_{\rm heat}(t')\,dt' -  \int_0^t t' L(t')\,dt' 
\end{align}
Eq. (\ref{eqn:integ_relation}) is similar to the analysis presented in \citep{Katz2013}, which considers times $t\gg t_{\rm peak}$ when $E(t)=0$.  Here we instead consider times around peak $t\sim t_{\rm peak}$. Furthermore, we assume the initial energy content in the ejecta is zero and ignore the initial stellar radius in our assumption of homology. Thus, the analysis presented here does not necessarily apply to Type IIP/L SNe, whose light curves are dominated by the initial shock-deposited energy.

We rewrite Eq. (\ref{eqn:integ_relation}) as
\begin{align}\label{eqn:integ_epsilon}
\frac{t^2}{2}L_{\rm peak} =  \int_0^t t' L_{\rm heat}(t')\,dt' +  \epsilon(t)
\end{align}
where
\begin{align}
\epsilon(t)= \left[ \frac{t^2}{2}L_{\rm peak}  -   \int_0^t t' L(t')\,dt'  \right] -  tE(t)
\end{align}
 The first term in brackets can be shown to be positive (since $L(t)\le L_{\rm peak}$) 
and monotonically increasing (see Appendix~B). The
second term $t E(t)$ is also positive and is a decreasing function when $L > L_{\rm heat}$, which
is typically obtained for $t \gtrsim t_{\rm peak}$. We therefore
anticipate there may be a time when the two functions cross and cancel to give $\epsilon(t)=0$.

We express this time as $t=\beta t_{\rm peak}$ and rearrange Eq.(\ref{eqn:integ_epsilon}) to get
\begin{align}\label{eqn:ptlr_eqn}
L_{\rm peak}=\frac{2}{\beta^2 t_{\rm peak}^2}\int_0^{\beta t_{\rm peak}}t' L_{\rm heat}(t')\,dt'
\end{align}
which is our desired expression for $L_{\rm peak}$.
 In Appendix B we show that  for common heating functions there is indeed a time when  $\epsilon(t)=0$ for a value of $\beta \sim 1$  that can be calibrated from the numerical simulations and is essentially independent of the heating timescale or  functional form.

Eq.(\ref{eqn:ptlr_eqn}) can be analytically evaluated for certain heating functions. For example, for an exponential heating function
\begin{align}
L_{\rm heat}(t)=L_0 e^{-t/t_s}
\end{align}
the peak time-luminosity relation can be evaluated to get
\begin{align}
L_{\rm peak}=\frac{2 L_0 t_s^2}{\beta^2 t_{\rm peak}^2}\left[1-(1+\beta t_{\rm peak}/t_s)e^{-\beta t_{\rm peak}/t_s}\right]
\label{eq:Lpeak_exp}
\end{align}
This can be contrasted with Arnett's rule, which predicts $L_{\rm peak} = L_0 e^{-t_{\rm peak}/t_s}$. The two expressions make  similar predictions 
when $t_{\rm peak} \ll t_s$ but increasingly diverges  for $t_{\rm peak} \gg t_s$.

\begin{figure}
    \centering
    \includegraphics[width=0.5\textwidth]{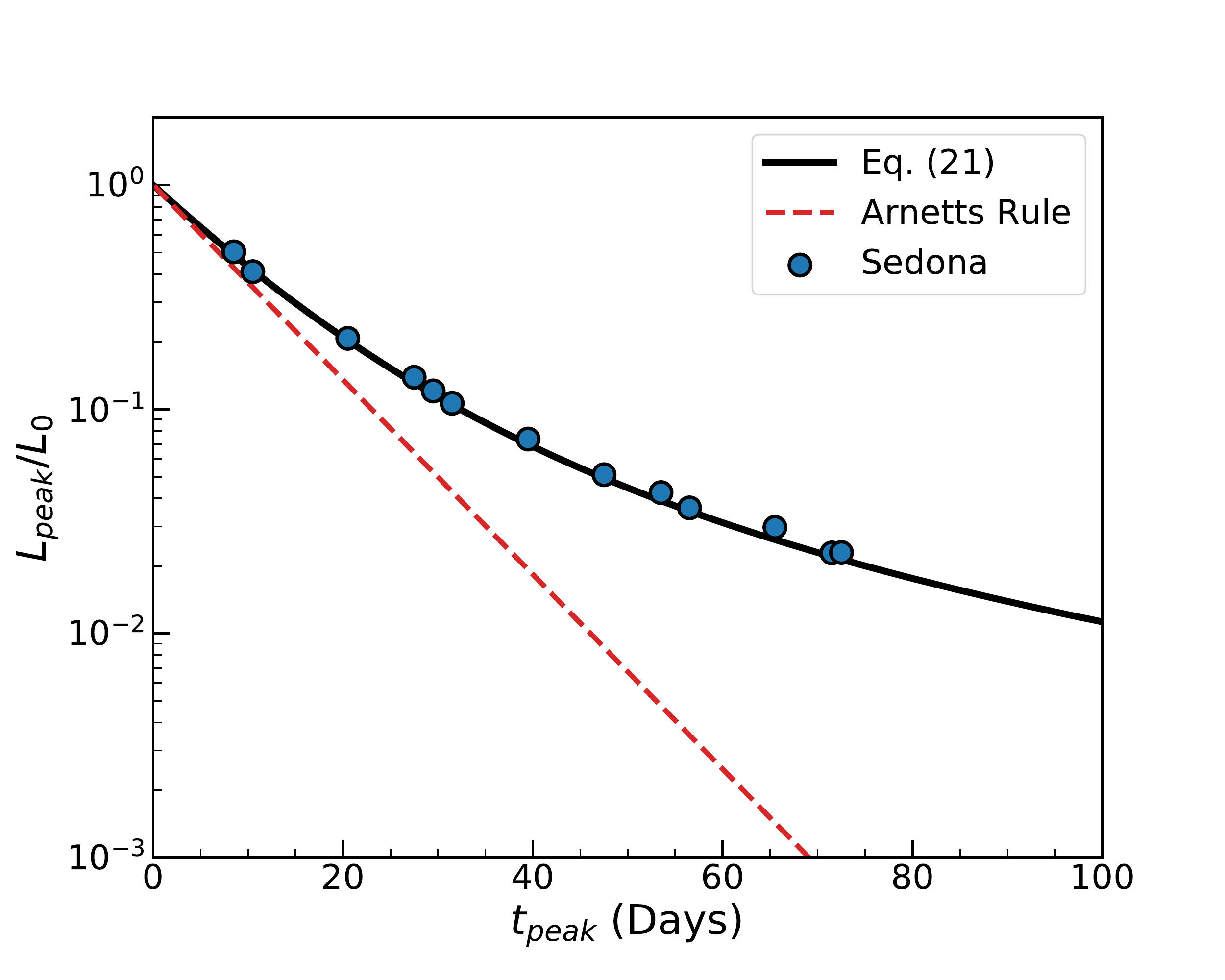}
    \caption{Relation between peak time and peak luminosity for a central exponential source with timescale $t_s=10$ days. Numerical radiation transport simulations with various peak times are shown (circles) compared to Arnett's rule (black dashed line) and the new relation Eq.(\ref{eqn:ptlr_eqn}) with $\beta=4/3$ (solid red line).}
    \label{fig:exp_10d_ptlr}
\end{figure}

In Fig. (\ref{fig:exp_10d_ptlr}) we compare our expression for $L_{\rm peak}$ to those of numerical light curve calculations for a central exponential heating source with $t_s=10$ days. The numerical models span a wide range of ejecta masses, velocities, and opacities, and hence result in a range of peak times.  Eq.~\ref{eq:Lpeak_exp} with $\beta=4/3$ gives a near-perfect match to the numerical simulations, independent of the ejecta properties. In comparison, Arnett's rule predicts systematically too low values of $L_{\rm peak}$ and becomes progressively worse for larger values of $t_{\rm peak}/t_s$.

\begin{figure}
    \includegraphics[width=0.5\textwidth]{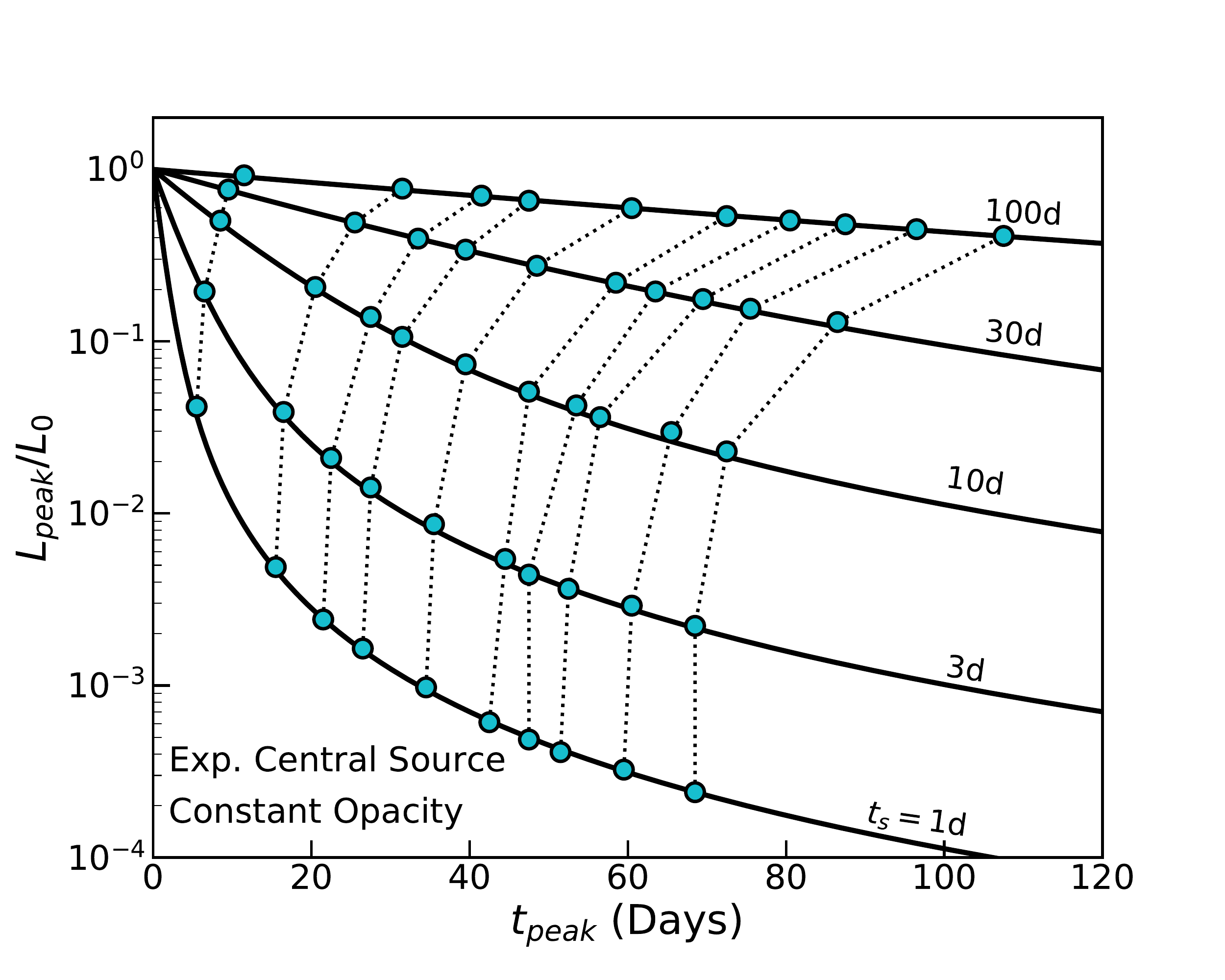}
    \includegraphics[width=0.5\textwidth]{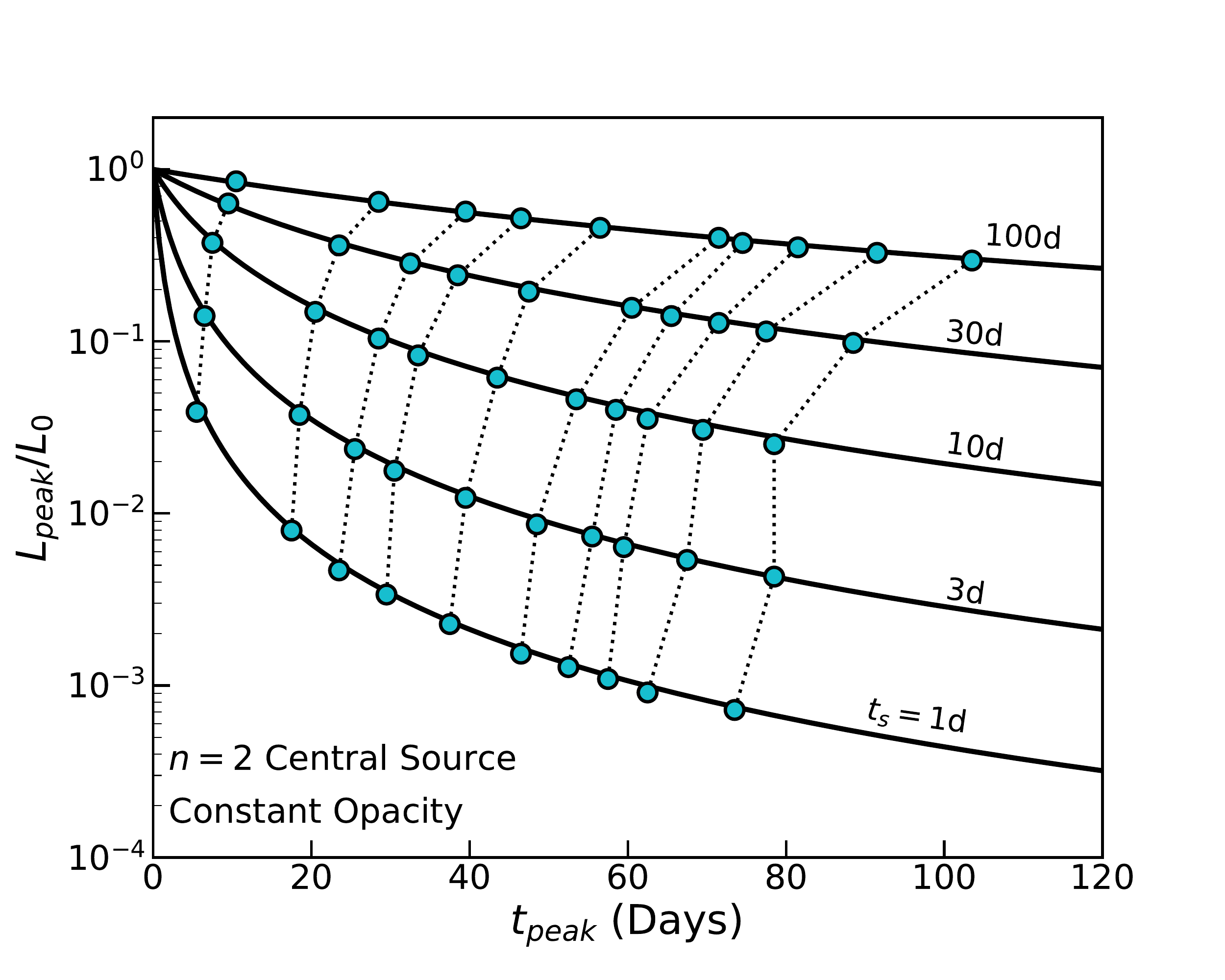}
    \caption{Same as Fig.(\ref{fig:exp_10d_ptlr}), but for different values of $t_s$, and for an exponential (top) and magnetar (bottom) source. Numerical simulations are shown as points. Lines of constant diffusion timescales $t_d$ are indicated (dotted black lines). Eq.(\ref{eqn:ptlr_eqn}) with $\beta=4/3$ is indicated by solid black lines, for a given $t_s$.}
    \label{fig:ptlr_exp_mag}
\end{figure}

The peak time-luminosity relation Eq.~\ref{eqn:ptlr_eqn} with $\beta=4/3$  applies for most central heating functions, as long as the opacity is constant and the density uniform. In Fig. \ref{fig:ptlr_exp_mag}, we show the peak time-luminosity relations for both an exponential source and a power-law source appropriate for magnetar energy injection
\begin{align}
L_{\rm heat}(t)=\frac{L_0}{(1+t/t_s)^2}
\end{align}
which can also be analytically evaluated (see Appendix~A). Fig. \ref{fig:ptlr_exp_mag} shows that the Eq.~\ref{eqn:ptlr_eqn} with $\beta=4/3$ accurately reproduces the numerical calculations. In later sections, we show that the value of $\beta$ does change if the source heating is spatially mixed or the opacity is non-constant due to recombination, but that $\beta$ remains largely independent of the heating function, source timescale $t_s$, or ejecta diffusion timescale $t_d$.

\section{Relation Between Peak Time and Diffusion Time}

\begin{figure}
    \centering
    \includegraphics[width=0.5\textwidth]{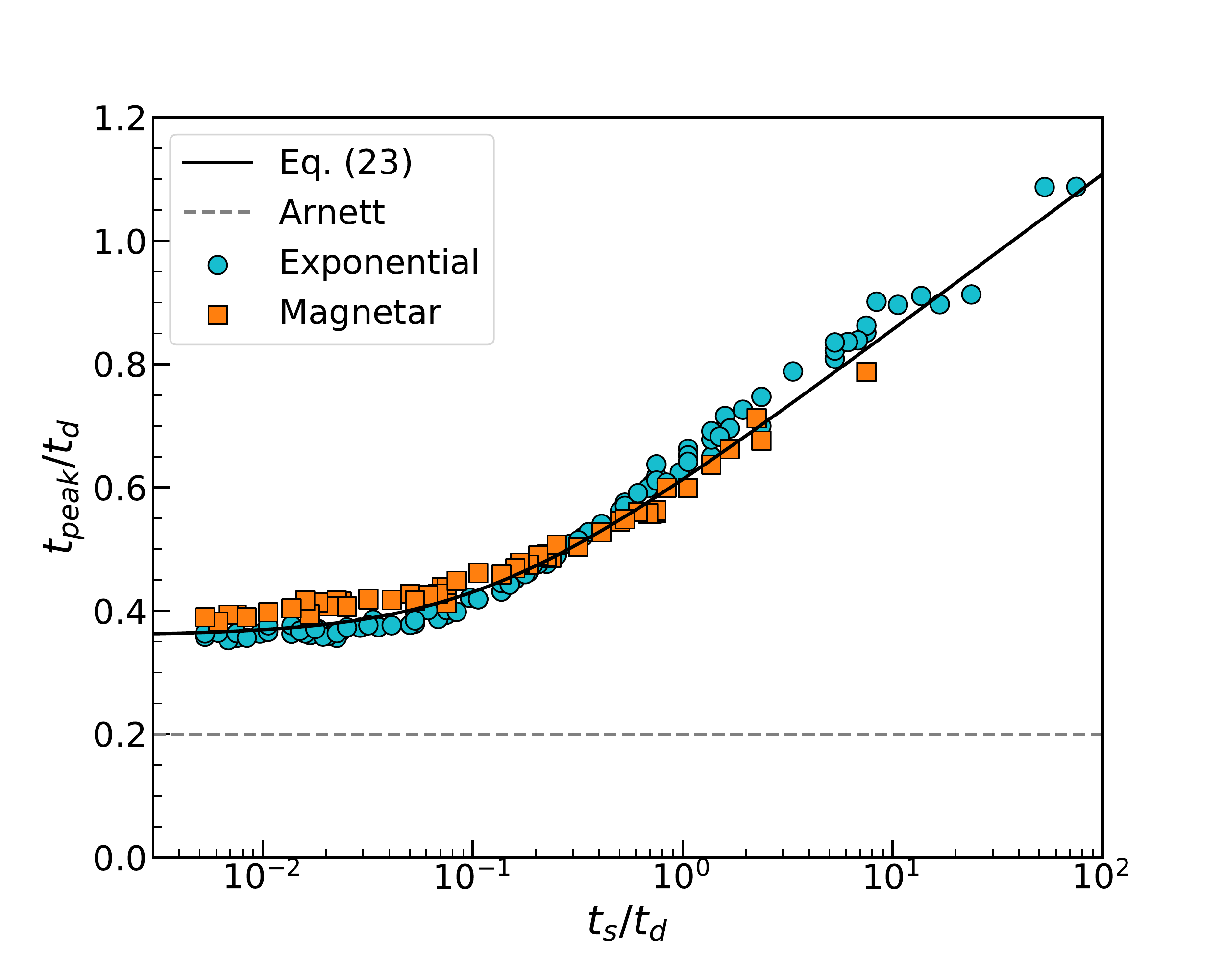}
    \caption{Relation between the source timescale $t_s$ and the peak time $t_{\rm peak}$, relative to the diffusion timescale $t_d$. Shown are an exponential (teal circles) and magnetar (orange squares) central heating source. The best-fit Eq.(\ref{eqn:ptlr_eqn}) is also shown (black dashed line). For comparison, the Arnett $t_{\rm peak}=t_d$ relation is shown.}
    \label{fig:tpeak_ts_td}
\end{figure}

Analyses of observed  light curves often attempt to constrain the ejecta mass and velocity by setting the observed time of peak, $t_{\rm peak}$, equal to the diffusion timescale $\tau_d$ \citep[e.g.,][]{Drout2011,Prentice2018}. Here we study that relation for constant opacity models, and show that $t_{\rm peak}$ depends not only on $t_d$, but also on the heating timescale $t_s$.

In Fig.(\ref{fig:tpeak_ts_td}), we show the dependence of $t_{\rm peak}$ on the ratio $t_s/t_d$, for a large number of numerical models with uniform density ejecta and two different central heating sources. The models have a range of masses, velocities, and constant opacities, although only the combination $t_d$ is relevant for the light curve behavior. For $t_s/t_d \ll 1$, the peak time asymptotes to $t_{\rm peak}\approx 0.4t_d$ independent of $t_s$. In this limit, the source can thus be approximated as an instantaneous ``pulse'' of energy deposited at $t_s$. The energy from such a pulse diffuses out and peaks at around $\sim 0.4 t_d$. In comparison, the Arnett models predict $t_{\rm peak}\approx 0.2t_d$ (see Fig. (\ref{fig:sedona_arnett_comp})).

For $t_s/t_d \gtrsim 0.1$, the continuing source deposition begins to lengthen the peak time.  The dependence is fairly weak -- $t_{\rm peak}$ only increases by a factor of $\sim 2$ as  $t_s$ changes over three orders of magnitude, implying that for the sources considered the light curve peak is mostly powered by heating deposited at early times.

An equation that captures the peak time of numerical models with constant opacity and central heating is
\begin{align}\label{eqn:tpeak}
\frac{t_{\rm peak}}{t_d}=0.11\ln\left(1+\frac{9t_s}{t_d}\right)+0.36
\end{align}
In the limit that $t_s\ll t_d$, Eq.(\ref{eqn:tpeak}) goes to $t_{\rm peak}\approx 0.4 t_d$, while for $t_s \gg t_d$ it grows
logarithmically with $t_s$. The relation is relatively insensitive to the functional form of the heating source (e.g. exponential vs. power-law) as long as the function is smoothly and gradually declining.

\section{Spatial Distribution of Heating}

\begin{figure}
    \centering
    \includegraphics[width=0.5\textwidth]{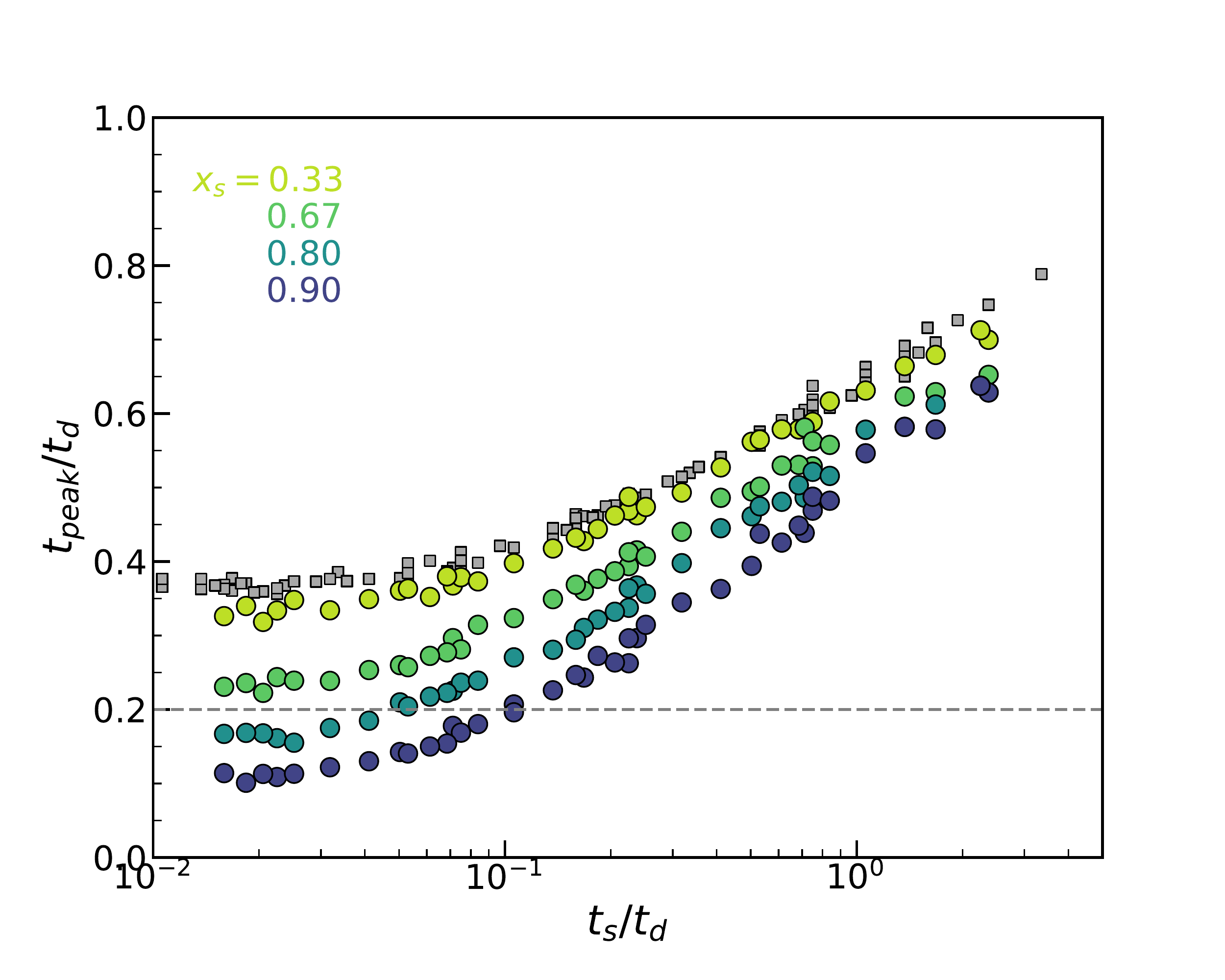}
    \caption{Relation between source timescale to the peak time, relative to the diffusion timescale. Different colors indicate different levels of concentration, parameterized by the concentration radius $x_s=0.33, 0.67, 0.8,$ and $0.90$. For comparison, the case of a centrally concentrated source ($x_s=0$) is shown (grey squares).}
    \label{fig:ts_tpeak_mix}
\end{figure}

\begin{figure}
    \centering
    \includegraphics[width=0.5\textwidth]{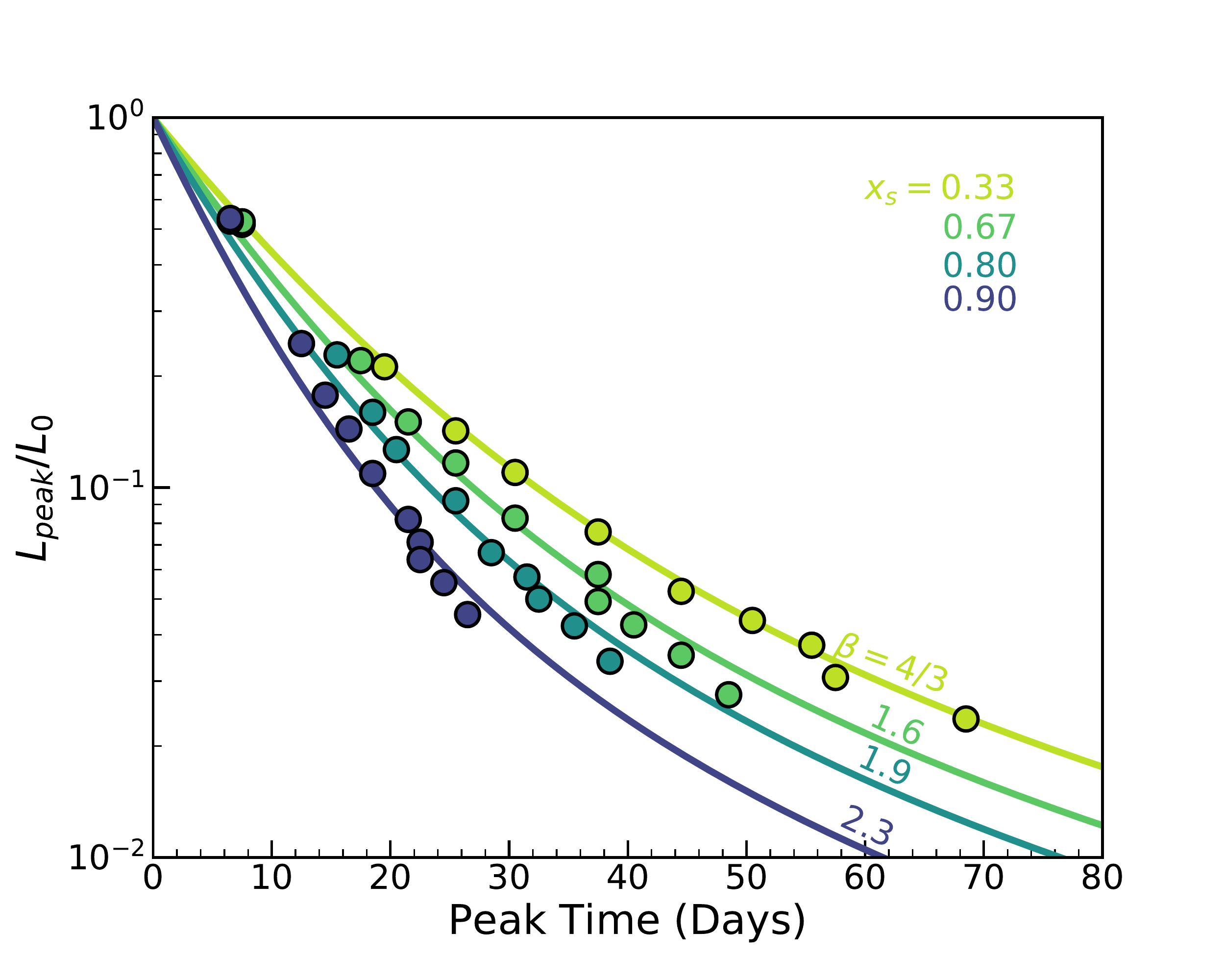}
    \caption{Peak time-luminosity relation for different spatial distributions of heating.}
    \label{fig:ptlr_mixing}
\end{figure}

Another important effect in shaping the light curve is the spatial distribution of heating  within the ejecta, e.g. different amounts of ``mixing'' of $^{56}$Ni in Type I SNe or assumed distribution of magnetar heating (see \cite{Dessart2018AIPTF14hls}). Indeed, In Fig. (\ref{fig:exp_mix_lcs}), we showed how the spatial distribution impacts both the peak time and luminosity of the light curve, and in particular found that the Arnett solution and Arnett's rule are  most appropriate for less concentrated/more uniform heating.

To account for the spatial distribution of heating, we take
the heating rate to be uniform out to a (scaled) radius $x_s$.  Fig.(\ref{fig:ts_tpeak_mix}) shows how the concentration affects the time of peak. 
The overall effect is to systematically drop the relation, i.e. for a given $t_s$ and $t_d$, concentration causes the light curve to peak earlier. This was shown for the case of $t_s$ = 10 days in Fig.(\ref{fig:exp_mix_lcs}). Interestingly, the peak time does not differ much unless the concentration radius is greater than $x_s>1/3$.

In all cases, there is similar behavior of a ``flattening'' in the relation for $t_s\ll t_d$. For the most mixed case $x_s=0.9$, the relation flattens as $t_{\rm peak}\approx 0.1t_d$. This lends further caution to using $t_d$ as a proxy for $t_{\rm peak}$; in addition to depending on $t_s$, there is also another dependence on $x_s$.

In Fig.(\ref{fig:ptlr_mixing}), we show the peak time-luminosity relation for the different spatial distributions of heating, for an exponential source with $t_s=10$ days. Interestingly, for different concentrations, the relation Eq.(\ref{eqn:ptlr_eqn}) still holds. The only difference is in the value of $\beta$. For $x_s=1/3$, the numerical simulations lie on the $\beta=4/3$ relation, which was found to be appropriate for a central source. This is in agreement with the results shown in Figs. (\ref{fig:exp_mix_lcs}) and (\ref{fig:ts_tpeak_mix}), where the $x_s=1/3$ does not differ significantly from simply assuming a central source. 

More centrally concentrated heating acts to increase the value of $\beta$. For the most uniform heating, $x_s=0.9$, $\beta$ increases by about a factor of $2$ compared to a central source. For the central exponential source used in Fig.(\ref{fig:ts_tpeak_mix}) and a constant opacity, we find that $\beta$ depends on $x_s$ approximately as
\begin{align}\label{eqn:xbeta}
\beta(x_s)\approx \frac{4}{3}\left(1+x_s^4\right)
\end{align}

Note that we assume local deposition of the heating source. In reality, for the case of e.g. $^{56}$Ni decay, there is the additional effect of gamma-ray deposition, which introduces a non-locality to the heating. In particular, gamma rays emitted closer to the center may deposit their energy farther out (or may escape entirely). Exploring this effect is outside the scope of this work (although see e.g. \cite{Dessart2016}).

\section{Non-Constant Opacity and Recombination}

\begin{figure}\label{fig:recomb_lc}
    \centering
    \includegraphics[width=0.5\textwidth]{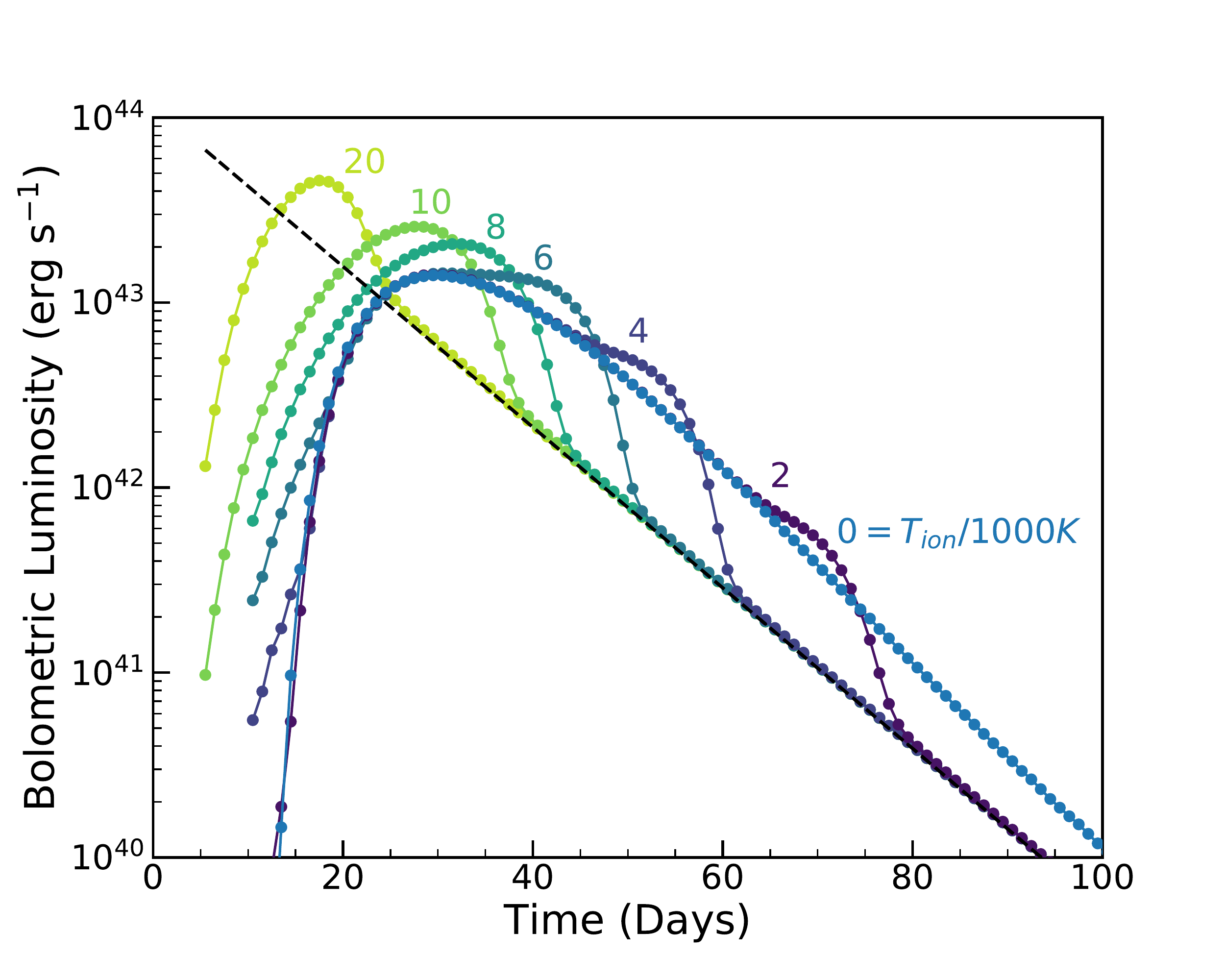}
    \caption{Light curves of a central exponential heating source with $t_s=10$ days and fixed ejecta properties, but varying the recombination temperature $T_{\rm ion}$. The input heating rate is shown (dashed black line). The case of $T_{\rm ion}=0$K is identical to assuming a constant grey opacity.}
    \label{fig:lc_tion_exp}
\end{figure}

While the previous results assumed a constant opacity, for certain compositions the opacity drops sharply when the ejecta cools and ions recombine. As the ejecta is typically hotter at the center, a cooling ``recombination front'' propagates from the surface inward \citep{Grassberg1971OnSupernovae,Grassberg1976TheShells,Popov1993}. 
 The photosphere is nearly coincident with the recombination front, with a temperature set by the ionization/recombination temperature $T_{\rm ion}$. 

To account for recombination effects in our numerical calculations, we prescribe a temperature dependence that mimics the behavior of the opacity in hydrogen and helium-rich compositions, for which  electron scattering dominates for $T>T_{\rm ion}$
\begin{align}
\kappa(T)=\kappa_0+\frac{\kappa_0-\epsilon\kappa_0}{2}\left[1+\tanh\left(\frac{T-T_{\rm ion}}{\Delta T_{\rm ion}}\right)\right].
\end{align}
The opacity $\kappa=\kappa_0$ for temperatures $T>T_{\rm ion}$ but drops to  $\kappa=\epsilon\kappa_0$ for $T<T_{\rm ion}$. The tanh function ensures a smooth transition over a temperature range $\Delta T_{\rm ion}$. We use $\epsilon=10^{-3}$ and $\Delta T_{\rm ion}=0.1T_{\rm ion}$, although our results are not sensitive to the exact values. We take the temperature $T$ to be equal to the radiation temperature $T_{\rm rad} = (E/a)^{1/4}$,
where $E$ is the energy density and $a$ the radiation constant. This is appropriate for a radiation-dominated ejecta and wavelength-independent opacity.

Fig.(\ref{fig:recomb_lc}) shows the effect of changing the recombination temperature $T_{\rm ion}$, while keeping  the heating source and ejecta properties fixed. These runs use a central exponential source with timescale $t_s=10$ days and energy $E_s=10^{50}$ ergs, and uniform ejecta with mass $M_{\rm ej}=5M_\odot$, velocity $v_{\rm ej}=10^9$ cm s$^{-1}$, and opacity $\kappa_0=0.1$ cm$^2$ g$^{-1}$.
For low $T_{\rm ion}$, most of the ejecta remains ionized at and after peak and the light curve resembles the constant opacity case, with the exception of a late-time ``bump" that occurs when recombination sets in and allows radiation to escape more easily. 
For higher $T_{\rm ion}$ recombination occurs earlier; for $T_{\rm ion} \gtrsim 6000$~K recombination results in a brighter and earlier light curve peak.

\begin{figure}\label{fig:recomb_mag}
    \centering
    \includegraphics[width=0.5\textwidth]{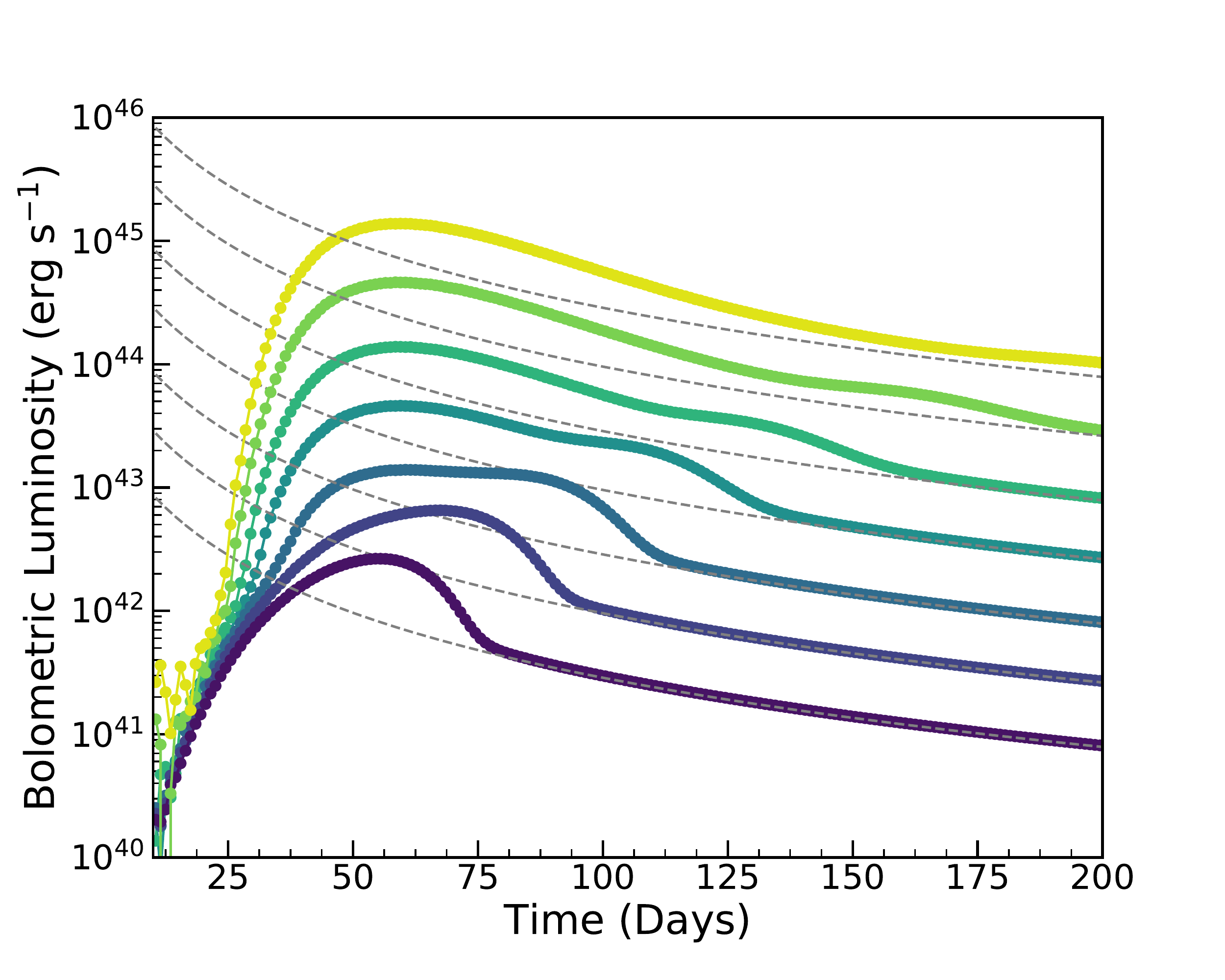}
    \caption{Light curves of a central magnetar heating source with $t_s=10$ days and fixed ejecta properties and recombination temperature $T_{\rm ion}=6000$K, but varying the heating energy $E_s$. The input heating rate for each light curve is shown (dashed grey).}
    \label{fig:lc_tion_es}
\end{figure}

When recombination is included, the total heating energy scale, $E_s$, impacts the  light curve morphology.  This is in contrast to constant opacity models, where $E_s$ simply sets the normalization of the light curve but leaves the shape the same. Fig. (\ref{fig:recomb_mag})  shows a set of  numerical light curves where only $E_s$ is varied.
At sufficiently large values of $E_s$, the heating source keeps the ejecta ionized until very late times, and  the light curve shape resembles a constant opacity light curve. As $E_s$ decreases, the ejecta recombines earlier, resulting in a ``bump'' at late times. For sufficiently low $E_s$, the heating source is unable to keep the ejecta temperature above $T_{\rm ion}$, and so recombination impacts the light curve peak.

\begin{figure}
    \centering
    \includegraphics[width=0.5\textwidth]{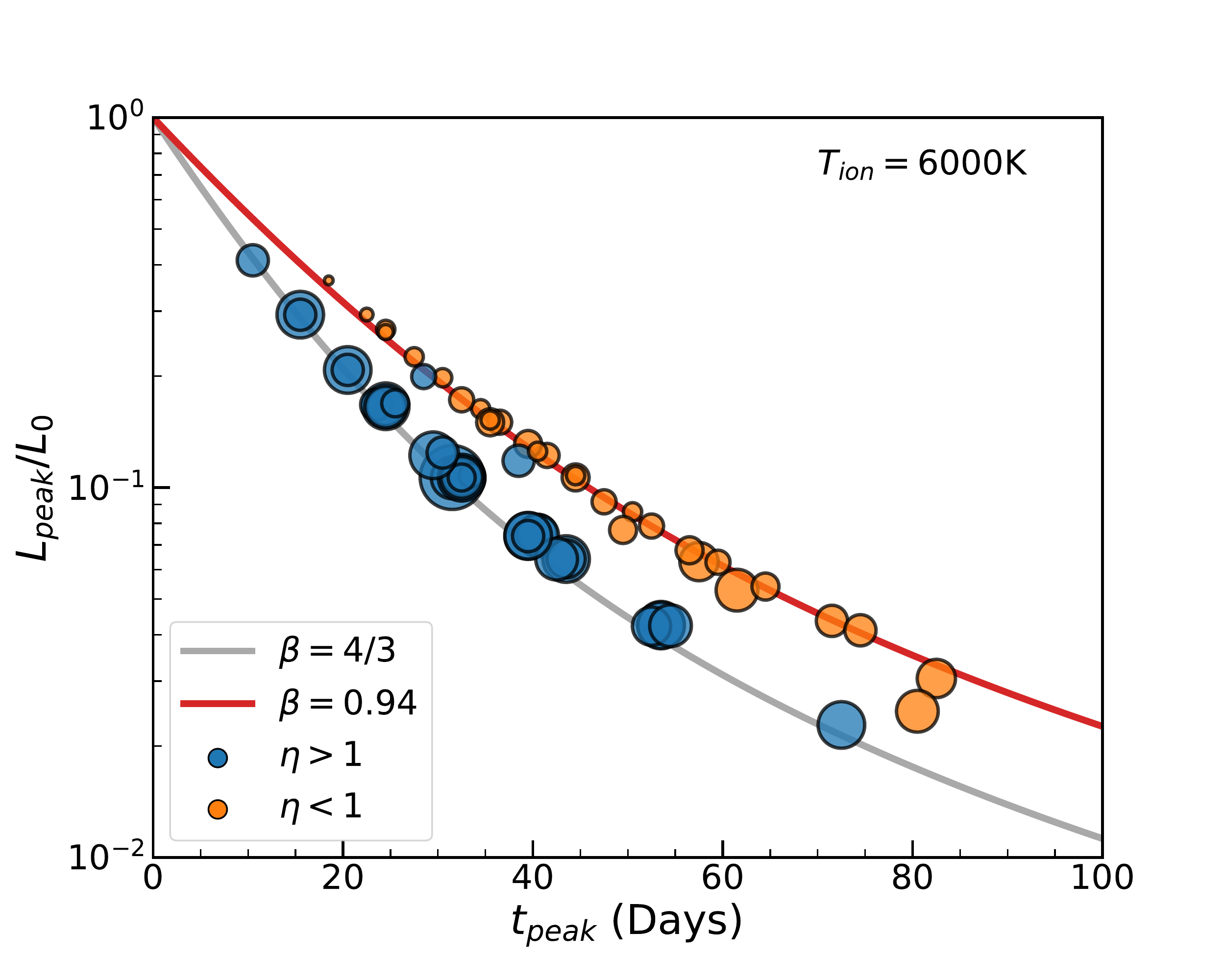}
    \caption{Peak time-luminosity relation with recombination effects, for a central exponential heating source with $t_s=10$ days. Points correspond to numerical simulations with $T_{\rm ion}=6000$K and  varying ejecta properties $M_{\rm ej}$, $v_{\rm ej}$, $\kappa_{\rm ej}$, as well different values of $E_s$, whose relative value is indicated by the point size. Blue and orange circles correspond to Eq.(\ref{eqn:eta}) with $\eta>1$ and $\eta<1$, respectively. Also shown is Eq.(\ref{eqn:ptlr_eqn}) for different values of $\beta$ (lines).}
    \label{fig:lpeak_tpeak_tion}
\end{figure}

To  determine whether recombination effects are important or not, we can compare the heating rate to the luminosity necessary to keep the ejecta ionized.  The ionizing luminosity is set by the ejecta radius and the recombination temperature as
\begin{align}
L_{\rm ion} \approx 4\pi R_{\rm ej}^2 \sigma_{\rm sb} T_{\rm ion}^4
\end{align}
From Section 3, the luminosity will roughly scale as
\begin{align}
L\sim \frac{E_s t_s}{t_d^2}
\end{align}
We define a ratio of the luminosity to the critical ionizing luminosity
\begin{align}
\eta &\equiv \frac{L}{L_{\rm ion}}
\propto \frac{c^2}{4\pi\sigma}\frac{E_s t_s}{\kappa^2 M_{\rm ej}^2 T_{\rm ion}^4}
\end{align}
We calibrate the proportionality based on numerical simulations of an exponential heating source to find
\begin{align}\label{eqn:eta}
\eta \sim 0.2\,E_{s,51}t_{s,10}\kappa_{0.2}^{-2}M_{\rm 10}^{-2}T_{4}^{-4}
\end{align}
where $E_{s,51}=E_s/10^{51}$ erg, $t_{s,10}=t_s/10$ days, $\kappa_{0.2}=\kappa/0.2$ cm$^2$ g$^{-1}$, $M_{10}=M_{\rm ej}/10M_\odot$, and $T_4=T_{\rm ion}/10^4$ K. For $\eta \lesssim 1$, the heating luminosity is too low to keep the ejecta sufficiently ionized and so recombination effects become important. 

In Fig.(\ref{fig:lpeak_tpeak_tion}), we show the results of a set of numerical simulations with $T_{\rm ion}=6000$K and various ejecta properties and $E_s$. Interestingly, the numerical simulations with recombination still fall on the relation Eq. (\ref{eqn:ptlr_eqn}). The only difference is that the value of $\beta$ changes. Specifically, recombination tends to \textit{decrease} the value of $\beta$. Also shown in Fig. (\ref{fig:lpeak_tpeak_tion}) are the respective values of $\eta$ for the numerical simulations. Points with $\eta>1$ are not affected by recombination at peak, and so fall on the relation with $\beta=4/3$, appropriate for a constant opacity. For $\eta<1$, recombination is important and the points fall on a smaller $\beta=0.94$ curve.

\section{Discussion and Conclusions}

\begin{figure}
    \centering
    \includegraphics[width=0.5\textwidth]{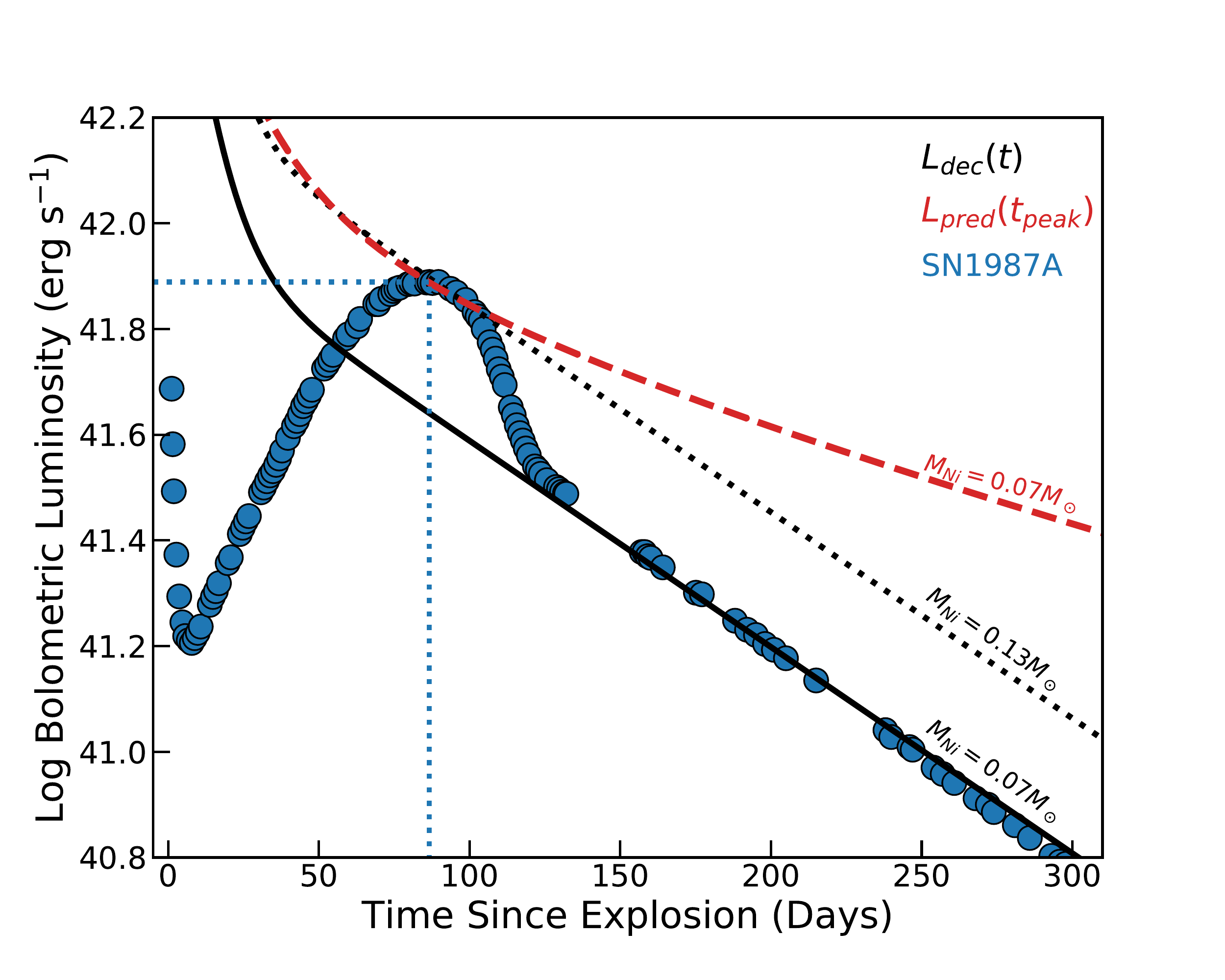}
    \caption{The light curve of SN1987A (blue points) from \cite{Suntzeff1990}, compared to input Ni+Co decay of $M_{\rm Ni}=0.07M_\odot$ (solid black) and $0.13M_\odot$ (dashed black). Also shown is Eq.(\ref{eqn:ptlr_eqn}) for $0.07M_\odot$ of Ni and $\beta=0.82$ (red dashed).}
    \label{fig:sn1987a}
\end{figure}

\begin{figure}
    \centering
    \includegraphics[width=0.5\textwidth]{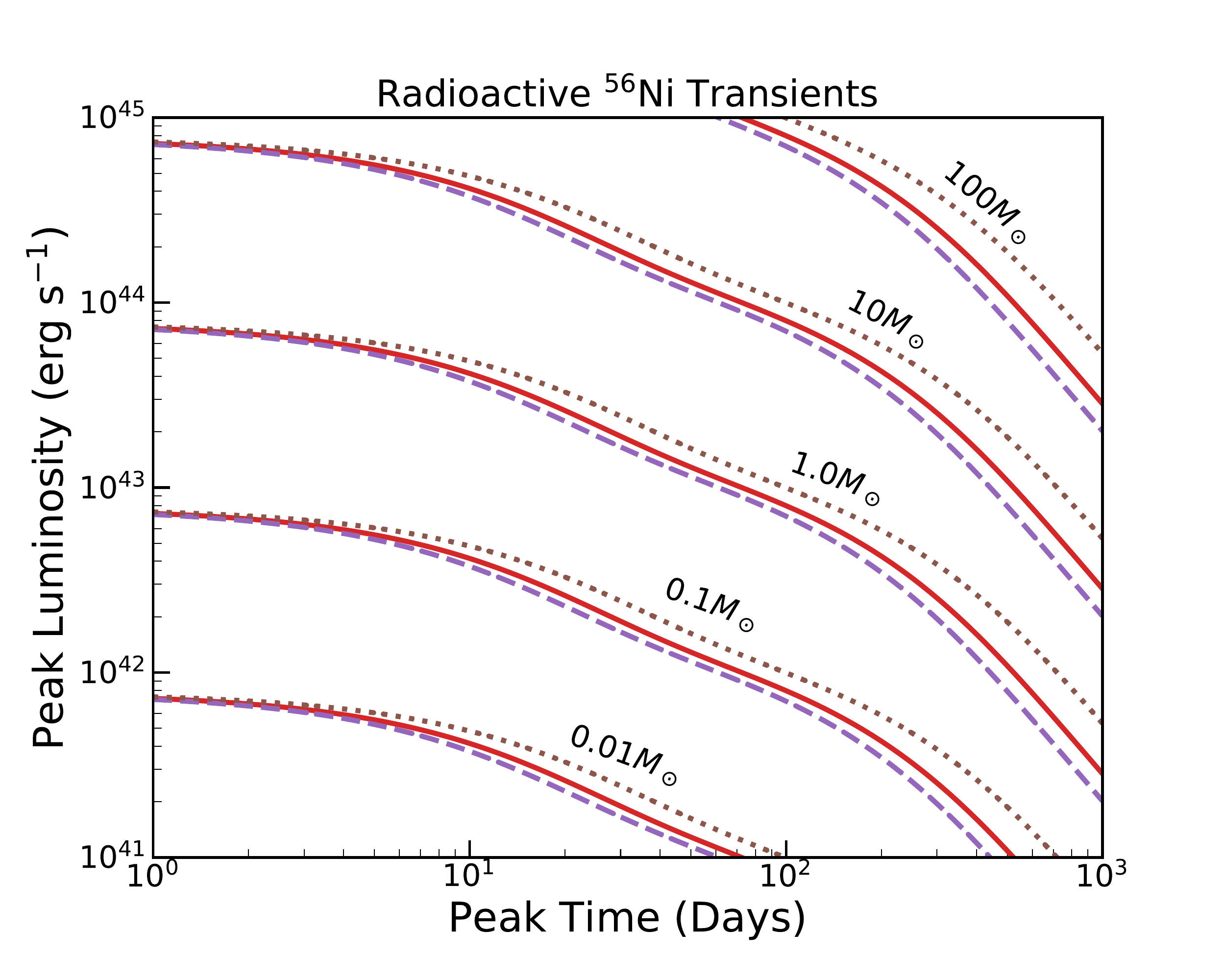}
    \caption{Peak time vs. peak luminosity for radioactive Ni-powered transients. Eq.(\ref{eqn:ptlr_eqn}) is shown for different values of $\beta=0.82$ (dotted brown), $\beta=9/8$ (solid red), and $\beta=4/3$ (dashed purple).}
    \label{fig:lpeak_tpeak_ni56}
\end{figure}

We have shown how the light curve peak time and luminosity are related for luminous transients, and derived analytic relations that can be used to infer the physical properties of the heating mechanism. In particular, Eq.(\ref{eqn:ptlr_eqn})
\begin{align*}
L_{\rm peak}=\frac{2}{\beta^2 t_{\rm peak}^2}\int_0^{\beta t_{\rm peak}}t' L_{\rm heat}(t')\,dt'
\end{align*}
captures the relationship between $t_{\rm peak}$ and $L_{\rm peak}$ where the light curve physics (i.e. recombination and concentration) is contained in the  $\beta$ parameter. Furthermore, Eq.(\ref{eqn:xbeta}) 
\begin{align*}
\beta(x_s)=\frac{4}{3}\left(1+x_s^4\right)
\end{align*}
gives the approximate dependence of $\beta$ on the spatial distribution of heating. Another useful result is given in Eq.(\ref{eqn:tpeak}) for central sources,

\begin{align*}
\frac{t_{\rm peak}}{t_d}=0.11\ln\left(1+\frac{9t_s}{t_d}\right)+0.36
\end{align*}
which shows how $t_{\rm peak}$ depends not only on the diffusion time $t_d$, but also the heating timescale $t_s$. In addition, recombination will change the value of $\beta$ compared to a constant opacity. In Table 1, we give approximate values of $\beta$ for a variety of transients. In Appendix A, we evaluate the peak time-luminosity relation for specific heating sources.

\begin{figure}
    \centering
    \includegraphics[width=0.5\textwidth]{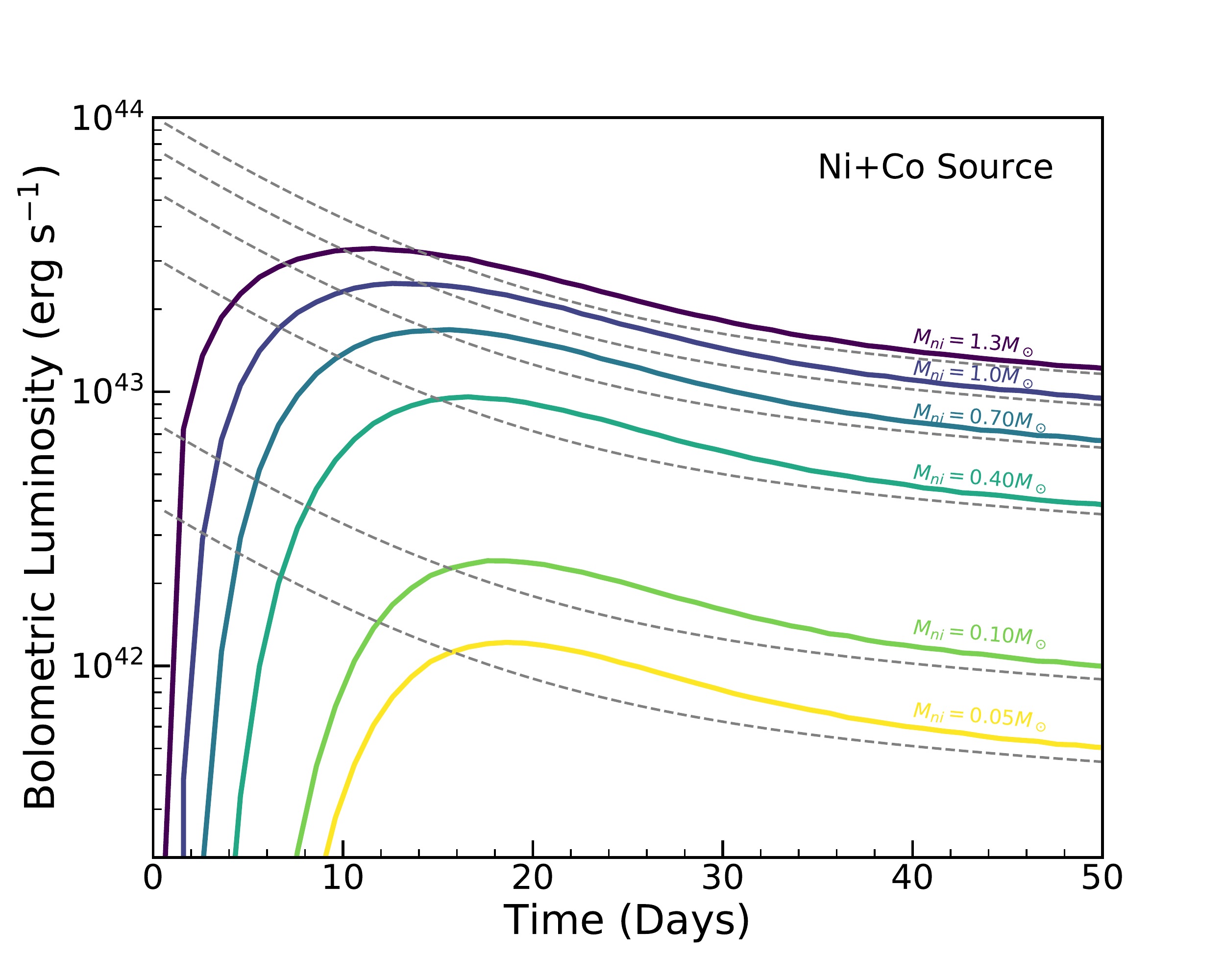}
    \caption{Bolometric light curves of toy Ia models, with $M_{\rm ej}=1.4M_\odot$, $v_{\rm ej}=10^9$ cm s$^{-1}$, $\kappa=0.1$ cm$^2$ g$^{-1}$ and assuming uniform density. Shown are light curves for different amounts of nickel masses $M_{\rm Ni}$ (heating rate shown as dashed grey lines) and, therefore, concentration.}
    \label{fig:nico_mix_lcs}
\end{figure}

For example, one of the common sources of heating in luminous transients is the radioactive decay chain of $^{56}$Ni \citep{Stritzinger2006,Valenti2007}. In particular, the decay chain of $^{56}$Ni $\rightarrow^{56}$Co $\rightarrow^{56}$Fe dominates the heating at timescales of interest. Fig.(\ref{fig:lpeak_tpeak_ni56}) shows the peak time-luminosity parameter space of $^{56}$Ni-powered transients, where the relation is given in Appendix A. For a given peak time and luminosity, one can thus infer an approximate value for the $^{56}$Ni mass for an appropriate choice in $\beta$.

We now consider the implications of our findings to analyzing observed SNe. In particular, Arnett's rule is often used to infer the nickel mass of Type Ia SNe. In Fig. (\ref{fig:nico_mix_lcs}) we show the effects of varying the nickel mass $M_{\rm Ni}$ in a set of toy Ia models with a total
 mass $M_{\rm ej} = 1.4 M_\odot$ and constant density and opacity. The inner layers of ejecta in these models are composed of pure $^{56}$Ni, and so higher $M_{\rm Ni}$ corresponds to a larger nickel core and less centrally concentrated heating. As expected, Arnett's rule works better for larger $M_{\rm Ni}$ and becomes progressively worse for the more centrally concentrated low $M_{\rm Ni}$ models. This suggests that analyses of SNe~Ia using
 Arnett's rule may be systematically biased, with the nickel mass of sub-luminous
 Ia's being overestimated.

As another case study,  we show in Fig. (\ref{fig:sn1987a}) the observed bolometric light curve of SN1987A, a Type II supernova whose primary peak is powered by radioactive $^{56}$Ni  \citep{Woosley1988,Suntzeff1990}. The late-time light curve behavior gives a constraint on the $^{56}$Ni mass to be $M_{\rm Ni}\approx 0.07M_\odot$. Arnett's rule  predicts a $M_{\rm Ni}$
a factor of $2$ too large, whereas using the new relation Eq.(\ref{eqn:ptlr_eqn}) with $\beta=0.82$ (appropriate for hydrogen recombination $T_{\rm ion}\approx 6000$K and a largely centrally-located Ni-Co heating source, as inferred from numerical simulations (see Fig.(\ref{fig:lpeak_tpeak_tion})), gives $M_{\rm Ni} \approx 0.07M_\odot$, in agreement with the late-time determination. 

\begin{figure}
    \centering
    \includegraphics[width=0.5\textwidth]{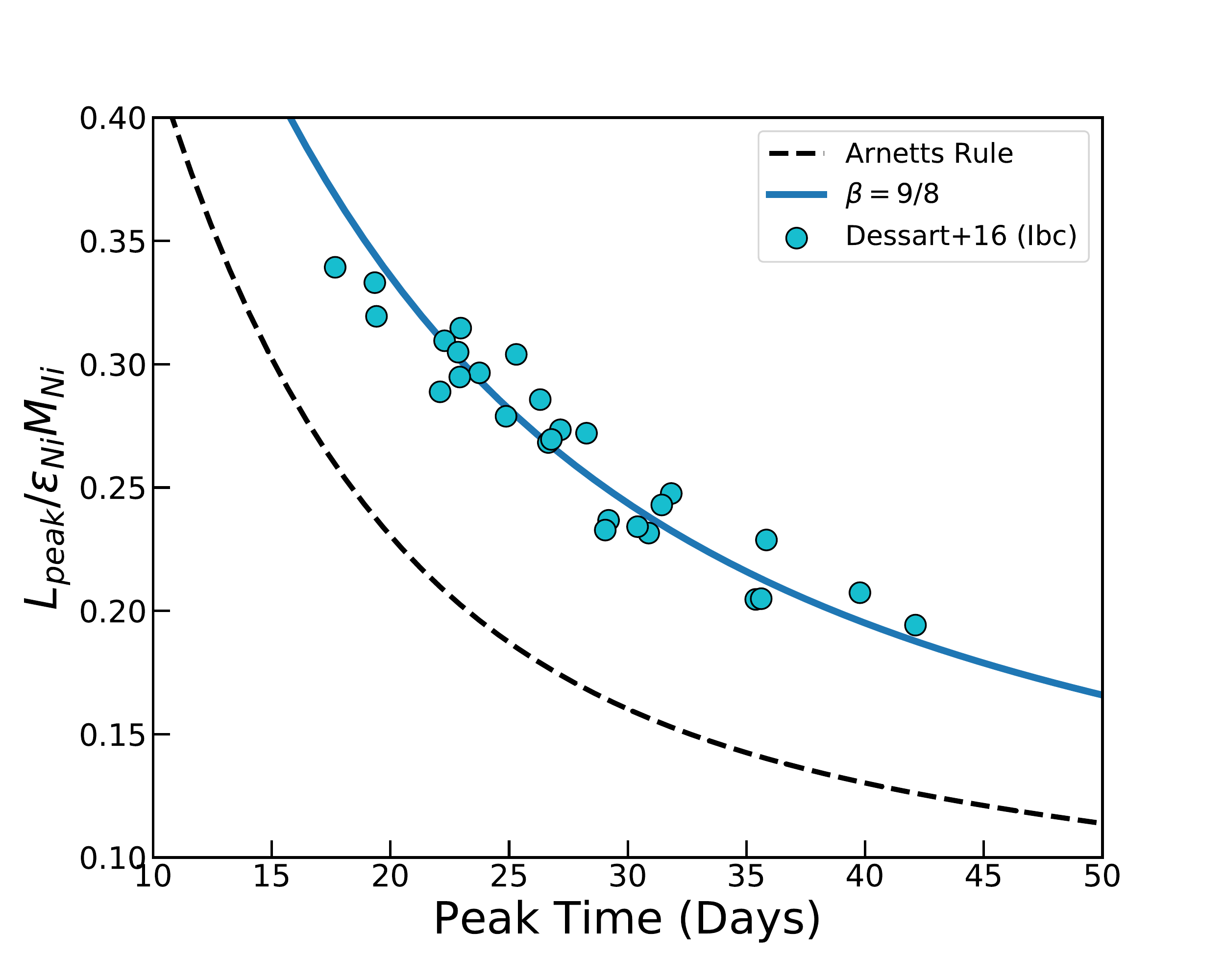}
    \caption{Peak time-luminosity relation of Eq.(\ref{eqn:ptlr_eqn}) compared to the Ibc models of \cite{Dessart2016}.}
    \label{fig:tpeak_lpeak_dessart}
\end{figure}

As another example,  we show in Fig.(\ref{fig:tpeak_lpeak_dessart}) the peak time-luminosity relation for the Type Ib/c SNe models presented in \cite{Dessart2016}. As noted in their work, Arnett's rule seems to overestimate the $^{56}$Ni mass of their models. Using Eq.(\ref{eqn:ptlr_eqn}), we find that the models lie on a $\beta=9/8$ relation. Given that the models do not have much mixing, we can assume centrally located heating and attribute any deviation from $\beta=4/3$ to recombination effects. Interestingly, a $\beta=9/8$ is in agreement with a recombination temperature of $T_{\rm ion}=4000$K, which is roughly that for a C- and O-rich composition. On the other hand, helium has a much higher recombination temperature and would imply a much smaller $\beta$; this indicates that the $^{56}$Ni in the \cite{Dessart2016} models are primarily diffusing out from the much denser carbon/oxygen inner ejecta rather than the outer helium ejecta. This is in agreement with the results in \cite{Piro2014TRANSPARENTSUPERNOVAE}, who similarly showed that light curve modeling is a better constraint on the C/O core rather than the helium.

The above examples demonstrate that, in principle, the peak time-luminosity relation may allow one to infer the composition of the ejecta \textit{solely from photometric observations}. Suppose we know from observations of the radioactive tail of SN1987A that it is powered by $0.07M_\odot$ of $^{56}$Ni,  and we assume the nickel to be largely centrally concentrated. From the peak time and luminosity we can solve for $\beta \approx 0.94$. Since each recombination temperature has its own unique value of $\beta$, we can then infer  $T_{\rm ion}\sim 6000$K, suggesting a hydrogen-rich composition.

If the composition, and hence recombination temperature, of  an observed supernova is constrained  (e.g. from spectroscopic  observations), then the derived value of $\beta$ may indicate the spatial distribution of the heating source. For example, one can assume to good approximation a constant opacity for Type~Ia SNe. Assuming central heating, this would point to a $\beta=4/3$, yet SNe-Ia seem to obey Arnett's rule fairly well, which
corresponds to a larger value of $\beta$ if using Eq.(\ref{eqn:ptlr_eqn}). This is in agreement with the results presented in Fig.(\ref{fig:ptlr_mixing}), since we expect Ia SNe to have a more uniform distribution of heating and hence fall on a larger $\beta$.

The main result, Eq.(\ref{eqn:ptlr_eqn}), is general enough to be applied to an arbitrary heating source, e.g. central-engine accretion, a magnetar, kilonovae, etc. Thus, for an observed transient peak time/luminosity that might be powered by other means than $^{56}$Ni, one must simply choose a different $L_{\rm in}(t)$ (which need not be analytic). Next, by choosing an appropriate $\beta$ (e.g. $4/3$ for constant opacity and a central source), one can constrain the heating source parameters. Note that there still remains a degeneracy in the heating source timescale $t_s$ and energy $E_s$. One is not able to break this degeneracy from the peak time and luminosity alone. Such constraints require additional information/observations or by putting physical limits on allowed values.

Several physical effects were neglected in our analysis here so as to isolate the basic behavior of supernova light curves. The models presented  assume spherical symmetry and adopt a grey opacity. Asymmetries in the ejecta/heating as well as non-grey effects likely play a role in the overall shape of the light curve, and on the inferred $\beta$ in the new relation. \cite{Dessart2018ImpactRadiation} show that clumping affects the recombination rate, which would impact the inferred $\beta$. We also used a simple parameterization for the spatial distribution of heating, which was taken to be uniform out to some radius; more complicated distributions (e.g. from accounting for gamma-ray deposition throughout the ejecta) warrant further investigation. 
We further made the assumption of homologous expansion, which may be violated
for supernovae interacting with a circumstellar material.
Because our relations apply to the bolometric peak, errors can also arise from observational effects, such as uncertainties in the  distance,  reddening,
or bolometric correction.


Conclusions drawn from the relation presented here are of course conditional on the  specific form of the heating source assumed (e.g. radioactive decay vs. magnetar spindown).
It is thus important to perform consistency checks on the nature of the heating source using additional information aside from the properties at peak, such as examining the slope of the late-time light curve tail.



While our comparisons here have demonstrated the limitations of the Arnett-like models, there exist other analytic models of transient light curves that attempt to account for the time-dependent diffusion effects that are important for setting the luminosity before and around peak  (e.g. \cite{Piro2013,Piro2014,Waxman2018ConstraintsEmission}). In future work, we will investigate a broader range of analytic models, and look for improved methods for calculating analytic light curves. Additionally, there exist other analytic techniques in estimating properties of the light curve, in particular the integral relation of \cite{Katz2013}. This requires knowing the full shape of the light curve out to times well after peak, rather than the properties at peak. The Katz integral approach and the new relation presented here are thus complementary in inferring the physical properties of the light curve.

There is still much work to be done to understand how/why the peak time-luminosity relation as well as it does, and in particular to better calibrate its value for specific heating mechanisms and ejecta properties. Nonetheless, the framework presented here will be useful for more detailed modeling, as well as providing a fast way to characterize the large number of transients to be discovered in current and upcoming surveys.

\acknowledgments
D.K.K was supported in part by the National Science Foundation Graduate Research Fellowship Program. D.K.K and D.N.K are supported by the U.S. Department of Energy, Office of Science, Office of Nuclear Physics, under contract number DE-AC02-05CH11231 and DE-SC0017616; SciDAC award DE-SC0018297; the Gordon and Betty Moore Foundation through Grant GBMF5076; and the Exascale Computing project (17-SC-20-SC), a collaborative effort of the U.S. Department of Energy and the National Nuclear Security Administration.

This research used resources of the National Energy Research Scientific Computing Center, a DOE Office of Science User Facility supported by the Office of Science of the U.S. Department of Energy under Contract No. DE-AC0205CH11231.
 
 \software{Sedona \citep{kasen2006}, Matplotlib, NumPy}

\bibliographystyle{aasjournal}
\bibliography{references}

\appendix

\numberwithin{equation}{section}

\begin{deluxetable*}{ccc}
\tablecaption{Peak Time-Luminosity Relation for Generic Heating Sources}
\tablehead{\colhead{Heating Type} & \colhead{Source Function $\mathcal{H}(t,t_s)$} & \colhead{$f(\tau,\beta)$}}
\startdata
Exponential & $\exp\left[-t/t_s\right]$ & $1-\left(1+\beta\tau\right)\exp\left[-\beta\tau\right]$ \\ 
Magnetar $n>2$ & $\frac{n-1}{(1+t/t_s)^n}$ & $\frac{1}{n-2}\left[1-\left(1+(n-1)\beta\tau\right)\left(1+\beta\tau\right)^{1-n}\right]$ \\
Magnetar $n=2$ & $\left(1+t/t_s\right)^{-n}$ & $\ln\left(1+\beta\tau\right)-\left(1+1/\beta\tau\right)^{-1}$ \\
Constant w/ Shutoff & $\Theta(t-t_s)$ & $1/2$  \\
Impulse & $E_0\delta(t-t_s)$ & $1$  \\
Power-Law $\alpha\ne 2$ & $\left(\frac{t}{t_s}\right)^{-\alpha}$ & $\frac{1}{2-\alpha}\left[(\beta\tau)^{2-\alpha}-1\right]$ \\
Power-Law $\alpha=2$ & $\left(\frac{t}{t_s}\right)^{-2}$ & $\ln\,\beta\tau$
\enddata
\end{deluxetable*}

\begin{deluxetable*}{cc}
\tablecaption{Values of $\beta$ for specific transients}
\tablehead{\colhead{Transient} & \colhead{$\beta$}}
\startdata
Generic - Central source, constant opacity & $4/3$ \\
Generic - Mixed source, constant opacity & $2$ \\
Generic - Central source, Hydrogen recombination & $0.94$ \\
Generic - Central source, Helium recombination & $0.7$ \\
Type Ia SNe & $1.6$ \\
Type Ibc SNe & $9/8$ \\
Type IIb/pec SNe & $0.82$
\enddata
\end{deluxetable*}

\section{Expressions for the Peak Time-Luminosity Relation}
From Section 3, we found an expression for the peak time-luminosity relation as
\begin{align}\label{eqn:app_lpeak}
L_{\rm peak}=\frac{2}{\beta^2 t_{\rm peak}^2}\int_0^{\beta t_{\rm peak}}\,t' L_{\rm heat}(t')\,dt'
\end{align}
where $L_{\rm peak}$ is the observed peak luminosity at the peak time $t_{\rm peak}$, $L_{\rm heat}(t)$ is the time-dependent heating rate, and $\beta$ is a constant that depends on opacity/concentration effects. Eq.(\ref{eqn:app_lpeak}) can be evaluated analytically for several functional forms of $L_{\rm heat}(t)$.

Suppose the heating source can be written most generally as
\begin{align}
L_{\rm heat}(t)= \frac{E_s}{t_s} \mathcal{H}(t,t_s)
\end{align}
where $\mathcal{H}(t,t_s)$ is the time-dependent component of $L_{\rm heat}(t)$, $t_s$ is the heating source timescale, and $E_s$ the characteristic heating energy. Equivalently, the heating source can be expressed in terms of a characteristic luminosity by setting $L_0=E_s/t_s$.

Let $\tau= t_{\rm peak}/t_s$ be the ratio between the peak time and source timescale. Then the peak luminosity can be evaluated to get
\begin{align}
L_{\rm peak}=\frac{2E_s t_s}{\beta^2 t_{\rm peak}^2}\times f(\tau,\beta)
\end{align}
where $f(\tau,\beta)$ depends on the exact functional form of $\mathcal{H}(t,t_s)$. For an exponential source,

\begin{align}
\mathcal{H}(t,t_s)=\exp\left[-t/t_s\right]
\end{align}

the integral can be evaluated to get
\begin{align}
f(\tau,\beta)=1-\left(1+\beta \tau\right)e^{-\beta \tau}
\end{align}
In Table 1, we give the analytic expressions of $f(\tau,\beta)$ for a variety of heating functions $\mathcal{H}(t,t_s)$. The choice of $\beta$ again depends on opacity/recombination effects, as well as the spatial distribution of heating. In Table 2, we give approximate values of $\beta$ based on numerical results.

\subsection{Radioactive Two-Decay Chain}
Numerous transients are powered by the radioactive decay of synthesized elements, e.g. $^{56}$Ni in Type I and IIb/pec supernovae.

Consider a decay chain consisting of $0\rightarrow 1\rightarrow 2$ with decay timescales $t_0$ and $t_1$, respectively (ignoring the contribution to heating of species 2).
The total number of the species at time $t$ is expressed as
\begin{align}
    &N_0(t)=Ne^{-t/t_0} \\
    &N_1(t) = N \frac{t_1}{t_1-t_0}\left(e^{-t/t_1}-e^{-t/t_0}\right)
\end{align}
where $N=N_0(0)$, and it is assumed that $N_1(0)=0$. Let $Q_0$, $m_0$, $Q_1$, and $m_1$ be the decay energies and species mass. Define heating rates per unit mass as
\begin{align}
    &\varepsilon_0 = \frac{Q_0}{m_0t_0} \\
    &\varepsilon_1 = \frac{Q_1}{m_1(t_1-t_0)}
\end{align}

Then the heating luminosity can be expressed as
\begin{align}
L_{\rm heat}(t)=M\left[(\varepsilon_0-\varepsilon_1)e^{-t/t_0}+\varepsilon_1 e^{-t/t_1}\right]
\end{align}
where $M=Nm_0$ is the total initial mass of species $0$.

With this expression we can derive the peak time-luminosity relation as
\begin{align}
\begin{split}
L_{\rm peak} = \frac{2\varepsilon_0 M t_0^2}{\beta^2 t_{\rm peak}^2}&\left[\left(1-\frac{\varepsilon_1}{\varepsilon_0}\right)\left(1-(1+\beta t_{\rm peak}/t_0)e^{-\beta t_{\rm peak}/t_0}\right)\right. \\
&\left. +\frac{\varepsilon_1 t_1^2}{\varepsilon_0 t_0^2}\left(1-(1+\beta t_{\rm peak}/t_1)e^{-\beta t_{\rm peak}/t_1}\right)\right]
\end{split}
\end{align}

\subsubsection{Radioactive Nickel Decay}

Supernovae of Type I and IIb/pec are powered primarily by the radioactive decay of $^{56}$Ni followed by $^{56}$Co \citep{Stritzinger2006,Valenti2007}. The heating function can be written in terms of the nickel mass $M_{\rm Ni}$ as

\begin{align}
L_{\rm heat}(t)=M_{\rm Ni}\left[\left(\varepsilon_{\rm Ni}-\varepsilon_{\rm Co}\right)e^{-t/t_{\rm Ni}}+\varepsilon_{\rm Co}e^{-t/t_{\rm Co}}\right]
\end{align}
where $\varepsilon_{\rm Ni}=3.9\cdot 10^{10}$ erg g$^{-1}$ s$^{-1}$ and $\varepsilon_{\rm Co}=6.8\cdot 10^9$ erg g$^{-1}$ s$^{-1}$ are the specific heating rates of Ni- and Co-decay, and $t_{\rm Ni}=8.8$ days and $t_{\rm Co}=111.3$ days are the decay timescales.

We can evaluate the peak time-luminosity relation in terms of $\beta$ as
\begin{align}
\begin{split}
L_{\rm peak}= \frac{2\varepsilon_{\rm Ni}M_{\rm Ni} t_{\rm Ni}^2}{\beta^2 t_{\rm peak}^2}& \left[\left(1-\frac{\varepsilon_{\rm Co}}{\varepsilon_{\rm Ni}}\right)\left(1-(1+\beta t_{\rm peak}/t_{\rm Ni})e^{-\beta t_{\rm peak}/t_{\rm Ni}}\right)\right.\\ 
&\left. +\frac{\varepsilon_{\rm Co}t_{\rm Co}^2}{\varepsilon_{\rm Ni}t_{\rm Ni}^2}\left(1-(1+\beta t_p/t_{\rm Co})e^{-\beta t_p/t_{\rm Co}}\right)\right]
\end{split}
\end{align}
Let $t_{p}=t_{\rm peak}/$day be the peak time in days. Using the numerical values of $\varepsilon_{\rm Ni}$, $\varepsilon_{\rm Co}$, $t_{\rm Ni}$,  and $t_{\rm Co}$ we get a more compact numerical expression (accurate to within $\sim 1\%$) as
\begin{align}
\begin{split}
L_{\rm peak} &=  10^{46}\left(\frac{M_{\rm Ni}}{M_\odot}\right)\,\frac{1}{\beta^2 t_p^2}\times \\
& \left[34.78-\left(1+0.114\beta t_p\right)\exp\left(-0.114\beta t_p\right)\right. \\
&\left. - 33.78\left(1+0.009\beta t_p)\exp\left(-0.009\beta t_p\right)\right)   \right]\,\,{\rm erg}\,\,{\rm s}^{-1}
\end{split}
\end{align}

%
%

\subsubsection{Magnetar-powered Supernovae}
The spindown luminosity of a magnetar is generally described
by \citep{Bodenheimer1974DoEvents,Gaffet1977PulsarSpectrum,Woosley2010BRIGHTBIRTH,Kasen2010}
\begin{align}
L_{\rm mag}(t)=\frac{E_{\rm mag}}{t_{\rm mag}}\frac{l-1}{\left(1+t/t_{\rm mag}\right)^l}
\end{align}
where $l=2$ for magnetic dipole spin-down,
\begin{align}
E_{\rm mag}=\frac{I_{\rm NS}\Omega^2}{2}=2\times 10^{50} P_{10}^{-2}\,\,{\rm erg}
\end{align}
is the magnetar energy with $P_{10}=P/10$ms is the spindown period, and
\begin{align}
t_{\rm mag}=\frac{6 I_{\rm NS}c^3}{B^2 R_{\rm NS}^6\Omega^2}=1.3B_{14}^{-2}P_{10}^2\,\,{\rm yr}
\end{align}
is the spindown timescale with $B_{14}=B/10^{14}$G the magnetic field strength. For $l=2$, we get
\begin{align}
L_{\rm peak}=\frac{2 E_{\rm mag} t_{\rm mag}}{\beta^2 t_{\rm peak}^2}\left[\ln(1+\beta\tau)-\left(1+1/(\beta\tau)\right)^{-1}\right]
\end{align}
were $\tau=t_{\rm peak}/t_{\rm mag}$.

\subsection{Accretion-Powered Transients}
Another interesting heating source is that of an accreting compact object \citep{Dexter2013}. Let $\dot{M}\sim M_{\rm acc}/t_{\rm acc}$ be the accretion rate of mass $M_{\rm acc}$ and timescale $t_{\rm acc}$. We here consider two functional forms of accretion luminosity. The first is of constant heating that ``shuts off'' after a time $t_{\rm acc}$,
\begin{align}
L_{\rm acc}(t)=\frac{\epsilon M_{\rm acc} c^2}{t_{\rm acc}}\Theta(t-t_{\rm acc})
\end{align}
where $\Theta(t-t_{\rm acc})$ is the Heaviside step function, and $\epsilon$ is the radiative efficiency. Substituting this into Eq.(\ref{eqn:ptlr_eqn}) we get
\begin{align}
L_{\rm peak}=\frac{\epsilon M_{\rm acc}t_{\rm acc}}{\beta^2 t_{\rm peak}^2}
\end{align}
Another functional form of interest is that of an $n=-5/3$ power law,
appropriate for fallback accretion
\begin{align}
L_{\rm acc}(t)=\frac{\epsilon M_{\rm acc}c^2}{t_{\rm acc}}\left(\frac{t}{t_{\rm acc}}\right)^{-5/3}
\end{align}
Again evaluating this source in Eq.(\ref{eqn:ptlr_eqn}) we get
\begin{align}
L_{\rm peak}=\frac{6\epsilon M_{\rm acc}c^2 t_{\rm acc}}{\beta^2 t_{\rm peak}^2}\left[\left(\frac{\beta t_{\rm peak}}{t_{\rm acc}}\right)^{1/3}-1\right]
\end{align}

\subsection{Kilonovae}
The kilonova heating rate from the radioactive decay of r-process elements can be parameterized as \citep{Li1998TransientMergers,Metzger2010ElectromagneticNuclei,Roberts2011ELECTROMAGNETICBINARIES,Lippuner2015R-PROCESSKILONOVAE}.

\begin{align}
L_{\rm heat}(t)=\varepsilon_0 M_{\rm ej}\left(\frac{t}{t_0}\right)^{-\eta}\times f(t)
\end{align}
where $\varepsilon_0\approx 10^{11}$ erg g$^{-1}$ s$^{-1}$ is the specific heating rate, $M_{\rm ej}$ is the ejecta mass, $t_0\approx 1$ day, and $\eta\approx 1.3$. The function $f(t)$ gives the thermalization efficiency with which the radioactive decay energy is able to deposit as heat in the ejecta.

\citep{Kasen2018} suggest an analytic approximation of the thermalization efficiency of electrons
\begin{align}
f(t)=\left(1+\frac{t}{t_e}\right)^{-1}
\end{align}
where
\begin{align}
t_e\approx 12.9 M_{0.01}^{2/3}v_{0.2}^{-2}\,\,{\rm days}
\end{align}
is the electron thermalization timescale, $M_{0.01}=M_{\rm ej}/0.01M_\odot$, and $v_{0.2}=v_{\rm ej}/0.2c$ the ejecta velocity.

Evaluating the integral for $\eta=1.3$ and the above approximation for the thermalization efficiency, we get
\begin{align}
L_{\rm peak}=2.86\varepsilon_0M_{\rm ej}\left(\frac{t_0}{\beta t_{\rm peak}}\right)^{1.3}\, _2F_1\left(0.7,1,1.7,-\beta t_{\rm peak}/t_e\right)
\end{align}
where $_2F_1(a,b,c,x)$ is the hypergeometric function which we approximate by a functional fit of
\begin{align}
_2F_1\left(0.7,1,1.7,-\beta t_{\rm peak}/t_e\right)\approx \left(1+\beta t_{\rm peak}/t_e\right)^{-1/2}
\end{align}
Thus, the peak time-luminosity relation for kilonova is approximately
\begin{align}
L_{\rm peak}=2.86\varepsilon_0M_{\rm ej}\left(\frac{t_0}{\beta t_{\rm peak}}\right)^{1.3}\cdot \left(1+\beta t_{\rm peak}/t_e\right)^{-1/2}
\end{align}
Assuming the r-process heating is uniformly mixed throughout the ejecta, we choose an approximate $\beta\approx 2$ based on numerical simulations.

\section{Derivation of the Peak Time-Luminosity Relation}

In Section 3, we showed that a simple relation holds between the peak time and luminosity of a light curve
\begin{align}
L_{\rm peak}=\frac{2}{\beta^2 t_{\rm peak}^2}\int_0^{\beta t_{\rm peak}} t' L_{\rm heat}(t')dt'
\end{align}
assuming there exists some time $t=\beta t_{\rm peak}$ such that $\epsilon(t)=0$, where we defined 
\begin{align}
\epsilon(t)=\left[\frac{t^2}{2}L_{\rm peak}-\int_0^t t' L(t')\,dt'\right]-tE(t)
\end{align}
Our goal, then, is to motivate that such a time when  $\epsilon(t)=0$ exists. We define two quantities
\begin{align}
\mathcal{F}(t)&= \frac{t^2}{2}L_{\rm peak}-\int_0^t t' L(t')\,dt'
\end{align}
and
\begin{align}
\mathcal{E}(t)=tE(t)
\end{align}
Thus, the time when $\epsilon(t)=0$ also implies
\begin{align}
\mathcal{F}(t)=\mathcal{E}(t)
\end{align}
Initially, $\mathcal{E}(t)$ will rise as heat is deposited and trapped in the optically thick ejecta. Eventually, the expanding ejecta becomes optically thin, allowing radiation to freely escape. As a result, any heating goes directly into the light curve, $L(t)=L_{\rm heat}(t)$. This implies that $\mathcal{E}(t)=0$ at late times $t\gg t_{\rm peak}$.

Since $\mathcal{E}(t)$ is continuous, if $\mathcal{F}(t)$ is a monotonically \textit{increasing} function of time and $\mathcal{F}(t)<\mathcal{E}(t)$ for some $t$, then it follows that $\mathcal{F}(t)$ and $\mathcal{E}(t)$ must intersect.

Taking the derivative in time of $\mathcal{F}(t)$ we get
\begin{align}
\mathcal{F}'(t)=t\left[L_{\rm peak}-L(t)\right]
\end{align}
Since $L(t)\leq L_{\rm peak}$ by definition, it follows that
\begin{align}
\mathcal{F}'(t)\geq 0
\end{align}
and so $\mathcal{F}(t)$ is indeed a monotonically increasing function of time.

Next, we need to show that there exists a time such that $\mathcal{F}(t)<\mathcal{E}(t)$. At $t=0$ we have $\mathcal{F}(t)=\mathcal{E}(t)=0$. For small $t$, we can expand the derivative of $\mathcal{E}(t)$ to get
\begin{align}
\mathcal{E}'(t)&\approx t\left[L_{\rm heat}(0)+tL'_{\rm heat}(0)-t^2L'(0)\right]\\
&\approx tL_{\rm heat}(0)+\mathcal{O}(t^2)
\end{align}
where we make use of the fact that $L(0)=0$. Similarly, for $\mathcal{F}(t)$, we get
\begin{align}
\mathcal{F}'(t)&\approx t\left[L_{\rm peak}-t^2 L'(0)\right]\\
&\approx tL_{\rm peak}+\mathcal{O}(t^3)
\end{align}
If the condition $L_{\rm heat}(0)>L_{\rm peak}$ is satisfied, then
\begin{align}
\mathcal{E}'(t)>\mathcal{F}'(t)
\end{align}
for small $t$. Since $\mathcal{F}(0)=\mathcal{E}(0)=0$, we have that, at early times, $\mathcal{E}(t)>\mathcal{F}(t)$. Combined with the monotonicity of $\mathcal{F}(t)$ and the fact that $\mathcal{E}(t)\rightarrow 0 $ at late times, it follows that $\mathcal{E}(t)$ and $\mathcal{F}(t)$ must intersect. In other words, there exists a time such that $\epsilon(t)=0$ and the peak time-luminosity relation holds.

There is a-priori mathematical justification for $L_{\rm heat}(0)>L_{\rm peak}$, however in physical cases of interest (e.g. radioactive decay, magnetar spindown, etc.), this seems to be a valid assumption, whereby $L_{\rm heat}(t)$ is monotonically decreasing in time. Furthermore, both diffusion and adiabatic degradation act to spread out and decrease the heating luminosity in time. This is confirmed in our numerical simulations for a wide variety of heating functions, wherein all the light curves seem to indicate $L_{\rm heat}(0)>L_{\rm peak}$ (see e.g. Figs. (\ref{fig:sedona_arnett_comp}) and \ref{fig:exp_mix_lcs}).

Another mathematical possibility is that $\mathcal{F}(t)$ and $\mathcal{E}(t)$ intersect more than once. Assuming $L_{\rm heat}(0)>L_{\rm peak}$, since $\mathcal{F}(t)$ is monotonically increasing, such behavior requires $\mathcal{E}(t)$ to ``oscillate'', i.e. there exists more than one time that $d\mathcal{E}/dt=0$. It is unclear whether such behavior is physical.

\begin{figure}
    \centering
    \includegraphics[width=0.5\textwidth]{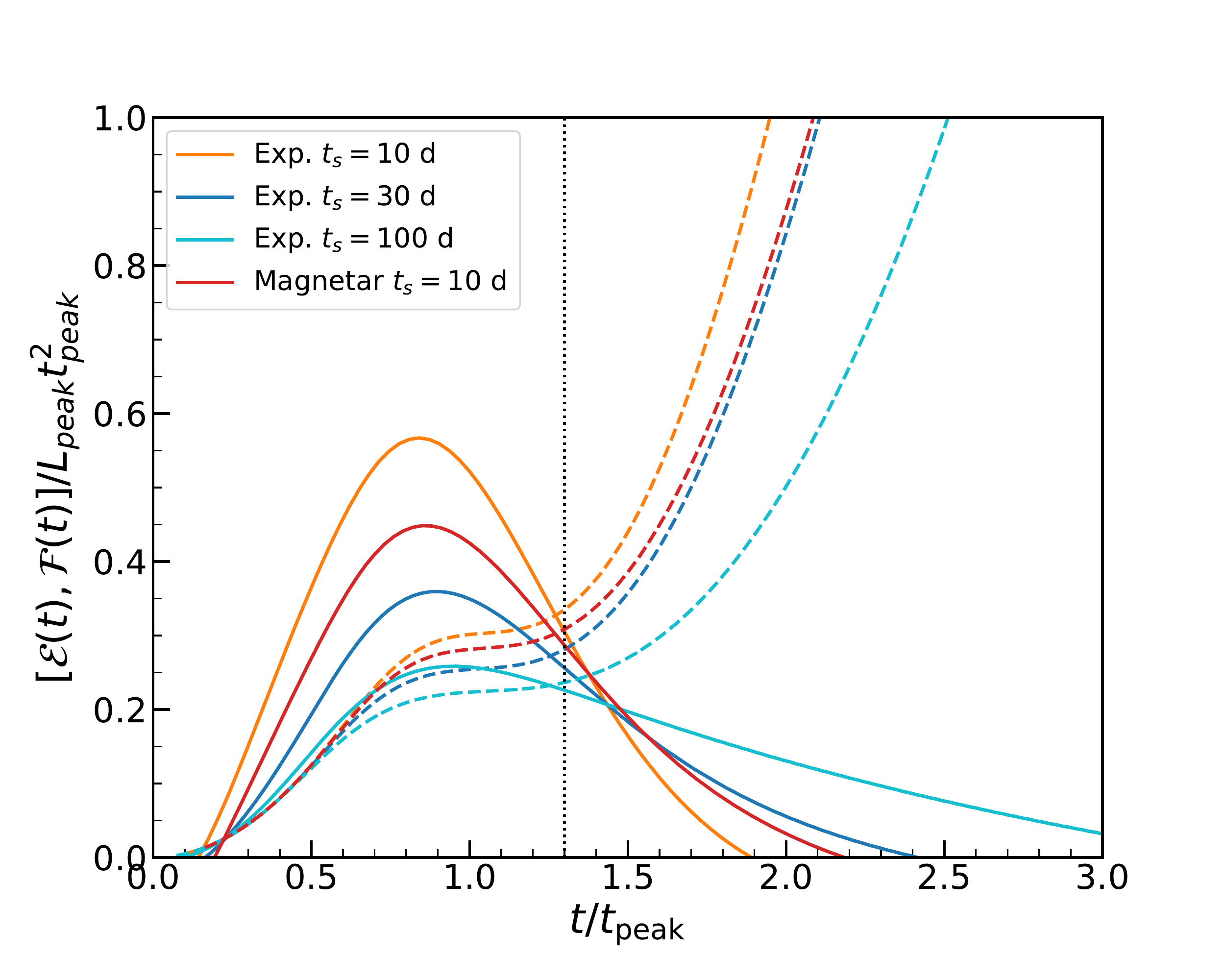}
    \caption{Comparison of the quantities $\mathcal{E}(t)$ (solid lines) and $\mathcal{F}(t)$ (dashed lines) as a function of time, for a central heating source with different functional forms and source timescales. The time when $\mathcal{E}(t)=\mathcal{F}(t)$ gives the value of $\beta$.}
    \label{fig:epsilon_plot}
\end{figure}

In summary, it is difficult to prove definitively that the peak time-luminosity relation holds for an arbitrary heating function. For a variety of monotonically decreasing heating functions, we have confirmed numerically that $L_{\rm heat}(0)>L_{\rm peak}$. Furthermore, this intersection appears to occur only once. In Fig.(\ref{fig:epsilon_plot}), we show the behavior of $\mathcal{E}(t)$ and $\mathcal{F}(t)$ for a subset of numerical simulations, assuming a central source and constant opacity. We pick different functional forms (exponential and magnetar-like) as well as different source timescales. In all cases, the time when $\epsilon(t)=0$ is nearly identical, with $\beta\approx 1.3$.

\end{document}